\documentclass[cp]{copernicus} % CP template

\usepackage{lscape}
\usepackage{longtable}
\usepackage{url}

\begin{document}

\title{HadISD: a quality-controlled global synoptic report database for selected variables at long-term stations from 1973--2011}

\author[1]{R.~J.~H.~Dunn}
\author[1]{K.~M.~Willett}
\author[2]{P.~W.~Thorne}
\author[*]{E.~V.~Woolley}
\author[3]{I.~Durre}
\author[4]{A.~Dai}
\author[1]{D.~E.~Parker}
\author[3]{R.~S.~Vose}

\affil[1]{Met Office Hadley Centre, FitzRoy Road, Exeter, EX1 3PB, UK}
\affil[2]{Cooperative Institute for Climate and Satellites, North Carolina State University and NOAA's National Climatic Data Center, Patton Avenue, Asheville, NC, 28801, USA}
\affil[3]{NOAA's National Climatic Data Center, Patton Avenue, Asheville, NC, 28801, USA}
\affil[4]{National Center for Atmospheric Research (NCAR), P.O. Box 3000, Boulder, CO 80307, USA}
\affil[*]{formerly at: Met Office Hadley Centre, FitzRoy Road, Exeter, EX1 3PB, UK}

\correspondence{R.~J.~H.~Dunn (robert.dunn@metoffice.gov.uk)}

\runningtitle{HadISD: a quality-controlled global synoptic database}
\runningauthor{R.~J.~H.~Dunn et al.}

\received{19 April 2012}
\pubdiscuss{21 May 2012}
\revised{24 September 2012}
\accepted{26 September 2012}

\firstpage{1649}

\published{25 October 2012}
\pubyear{2012}
\pubvol{8}
\pubnum{5}
\maketitle

\begin{abstract}
This paper describes the creation of HadISD: an automatically
quality-controlled synoptic resolution dataset of temperature,
dewpoint temperature, sea-level pressure, wind speed, wind direction
and cloud cover from global weather stations for 1973--2011. The full
dataset consists of over 6000 stations, with 3427 long-term stations
deemed to have sufficient sampling and quality for climate
applications requiring sub-daily resolution. As with other surface
datasets, coverage is heavily skewed towards Northern \mbox{Hemisphere}
mid-latitudes.

The dataset is constructed from a large pre-existing ASCII flatfile
data bank that represents over a decade of substantial effort at data
retrieval, reformatting and provision.  These raw data have had
varying levels of quality control applied to them by individual data
providers. The work proceeded in several steps: merging stations with
multiple reporting identifiers; reformatting to netCDF; quality
control; and then filtering to form a final dataset. Particular
attention has been paid to maintaining true extreme values where
possible within an automated, objective process. Detailed validation
has been performed on a subset of global stations and also on UK data
using known extreme events to help finalise the QC tests.  Further
validation was performed on a selection of extreme events world-wide
(Hurricane Katrina in 2005, the cold snap in Alaska in 1989 and heat
waves in SE Australia in 2009). Some very initial analyses are
performed to illustrate some of the types of problems to which the
final data could be applied. Although the filtering has removed the
poorest station records, no attempt has been made to homogenise the
data thus far, due to the complexity of retaining the true
distribution of high-resolution data when applying adjustments. Hence
non-climatic, time-varying errors may still exist in many of the
individual station records and care is needed in inferring long-term
trends from these data.

This dataset will allow the study of high frequency variations of
temperature, pressure and humidity on a global basis over the last
four decades.  Both individual extremes and the overall population of
extreme events could be investigated in detail to allow for comparison
with past and projected climate.  A version-control system has been
constructed for this dataset to allow for the clear documentation of
any updates and corrections in the future.
\end{abstract}

\introduction\label{sec:intro}

The Integrated Surface Database (ISD) held at NOAA's National Climatic
Data Center is an archive of synoptic reports from a large number of
global surface stations (\citealp{Smith11, Lott04}; see
\url{http://www.ncdc.noaa.gov/oa/climate/isd/index.php}). It is a rich
source of data useful for the study of climate variations, individual
meteorological events and historical climate impacts. For example,
these data have been applied to quantify precipitation frequency
\citep{Dai01a} and its diurnal cycle \citep{Dai01b}, diurnal variations
in
surface winds and divergence field \citep{Dai99}, and recent
changes in surface humidity \citep{Dai06a, Willett08},
cloudiness \citep{Dai06b} and wind speed \mbox{\citep{Peterson11}.}

The collation of ISD, merging and reformatting to a single format from
over 100 constituent sources and three major databanks represented a
substantial and ground-breaking effort undertaken over more than a
decade at NOAA NCDC. The database is updated in near real-time. A
number of automated quality control (QC) tests are applied to the data
that largely consider internal station series consistency and are
geographically invariant in their application (i.e.~threshold values
are the same for all stations regardless of the local
climatology). These procedures are briefly outlined in \citet{Lott04} and
\citep{Smith11}. The tests concentrate on the most widely used
variables and consist of a mix of logical consistency checks and
outlier type checks. Values are flagged rather than deleted. Automated
checks are essential as it is impractical to manually check thousands
of individual station records that could each consist of several tens
of thousands of individual observations. It should be noted that the
raw data in many cases have been previously quality controlled
manually by the data providers, so the raw data are not necessarily
completely raw for all stations.

The ISD database is non-trivial for the non-expert to access and use,
as each station consists of a series of annual ASCII flatfiles (with
each year being a separate directory) with each observation
representing a row in a format akin to the synoptic reporting codes
that is not immediately intuitive or amenable to easy machine reading
(\url{http://www1.ncdc.noaa.gov/pub/data/ish/ish-format-document.pdf}).
NCDC, however, provides access to the ISD database using a GIS interface.
This does give the ability for users to select parameters and stations
and output the results to a text file.  Also, a subset of the ISD
variables (air temperature, dewpoint temperature, sea level pressure,
wind direction, wind speed, total cloud cover, one-hour accumulated
liquid precipitation, six-hour accumulated liquid precipitation) is
available as ISD-Lite in fixed-width format ASCII files.  However,
there has been no selection on data or station quality.
In this paper we outline the steps undertaken to provide a new quality-controlled version,
called HadISD, which is based on the raw ISD
records, in {netCDF} format for selected variables for a subset of the
stations with long records.   This new dataset will allow the easy
study of the behaviour of short-timescale climate phenomena in recent
decades, with the subsequent comparison to past climates and future
climate projections.

One of the primary uses of a sub-daily resolution database will be the
characterisation of extreme events for specific locations, and so it
is imperative that multiple, independent efforts be undertaken to
assess the fundamental quality of individual observations. We also
therefore undertake a new and comprehensive quality control of the
ISD, based upon the raw holdings, which should be seen as
complementary to that which already exists. In the same way that
multiple independent homogenisation efforts have informed our
understanding of true long-term trends in variables such as
tropospheric temperatures (Thorne et al., 2011), numerous independent
QC efforts will be required to fully understand changes in
extremes. Arguably, in this context structural uncertainty
\citep{Thorne05} in quality control choices will be as important as that in
any homogenisation processes that were to be applied in ensuring an
adequate portrayal of our true degree of uncertainty in extremes
behaviour. Poorly applied quality control processes could certainly
have a more detrimental effect than poor homogenisation processes. Too
aggressive and the real tails are removed; too liberal and data
artefacts remain to be misinterpreted by the unwary.  As we are
unable to know for certain whether a given value is truly valid, it is
impossible to unambiguously determine the prevalence of type-I and
type-II errors for any candidate QC algorithm.  In this work, type-I
errors occur when a good value is flagged, and type-II errors are when
a bad value is not flagged.

Quality control is therefore an increasingly important aspect of
climate dataset construction as the focus moves towards regional- and
local-scale impacts and mitigation in support of climate services
\citep{Doherty08}. The data required to support these
applications need to be at a much finer temporal and spatial
resolution than is typically the case for most climate datasets, free
of gross errors and homogenised in such a way as to retain the high as
well as low temporal frequency characteristics of the
record. Homogenisation at the individual observation level is a
separate and arguably substantially more complex challenge. Here we
describe solely the data preparation and QC. The methodology is
loosely based upon that developed in \citet{Durre10} for daily
data from the Global Historical Climatology Network. Further
discussion of the data QC problem, previous efforts and references can
be found therein. These historical issues are not covered in any
detail here.

Section~\ref{sec:compositing} describes how stations that report under varying identifiers
were combined, an issue that was found to be globally insidious and
particularly prevalent in certain regions. Section~\ref{sec:selection} outlines
selection of an initial set of stations for subsequent QC. Section \ref{sec:QC}
outlines the intra- and inter-station QC procedures developed and
summarises their impact. We validate the final quality-controlled
dataset in Sect.~\ref{sec:validation}.  Section~\ref{sec:finalselection} briefly summarises the final
selection of stations, and Sect.~\ref{sec:nomenclature} describes our
version numbering system. Section~\ref{sec:uses} outlines some very simple analyses of
the data to illustrate their likely utility, whilst Sect.~\ref{sec:summary}
concludes.

The final data are available through
\url{http://www.metoffice.gov.uk/hadobs/hadisd} along with the large volume
of process metadata that cannot reasonably be appended to this
paper. The database covers 1973 to end-2011, because availability
drops off substantially prior to 1973 \citep{Willett08}. In
future periodic updates are planned to keep the dataset \mbox{up-to-date.}

\begin{figure}[t]
\includegraphics[width=8.5cm]{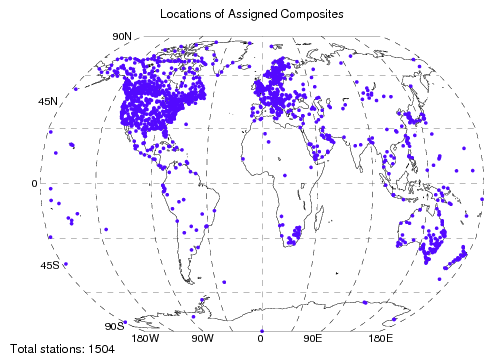}
\includegraphics[width=8.5cm]{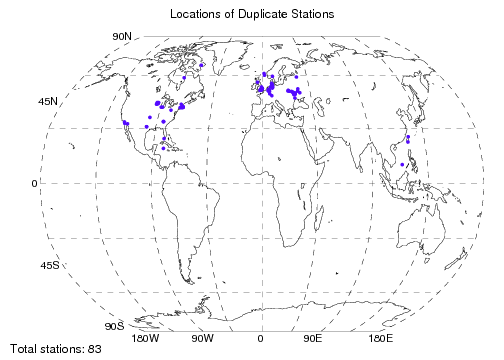}
\caption{\label{fig:1}Top:~locations of assigned composite stations from
  the ISD database before any station selection and filtering.  Only
  943 of these 1504 stations were passed into the QC process.
  Bottom:~locations of 83 duplicated stations identified by the inter-station
  duplicate check (Sect.~\ref{sec:QC:tests:1}, test 1.) }
\end{figure}

\begin{figure}[p]
\includegraphics[width=8.5cm]{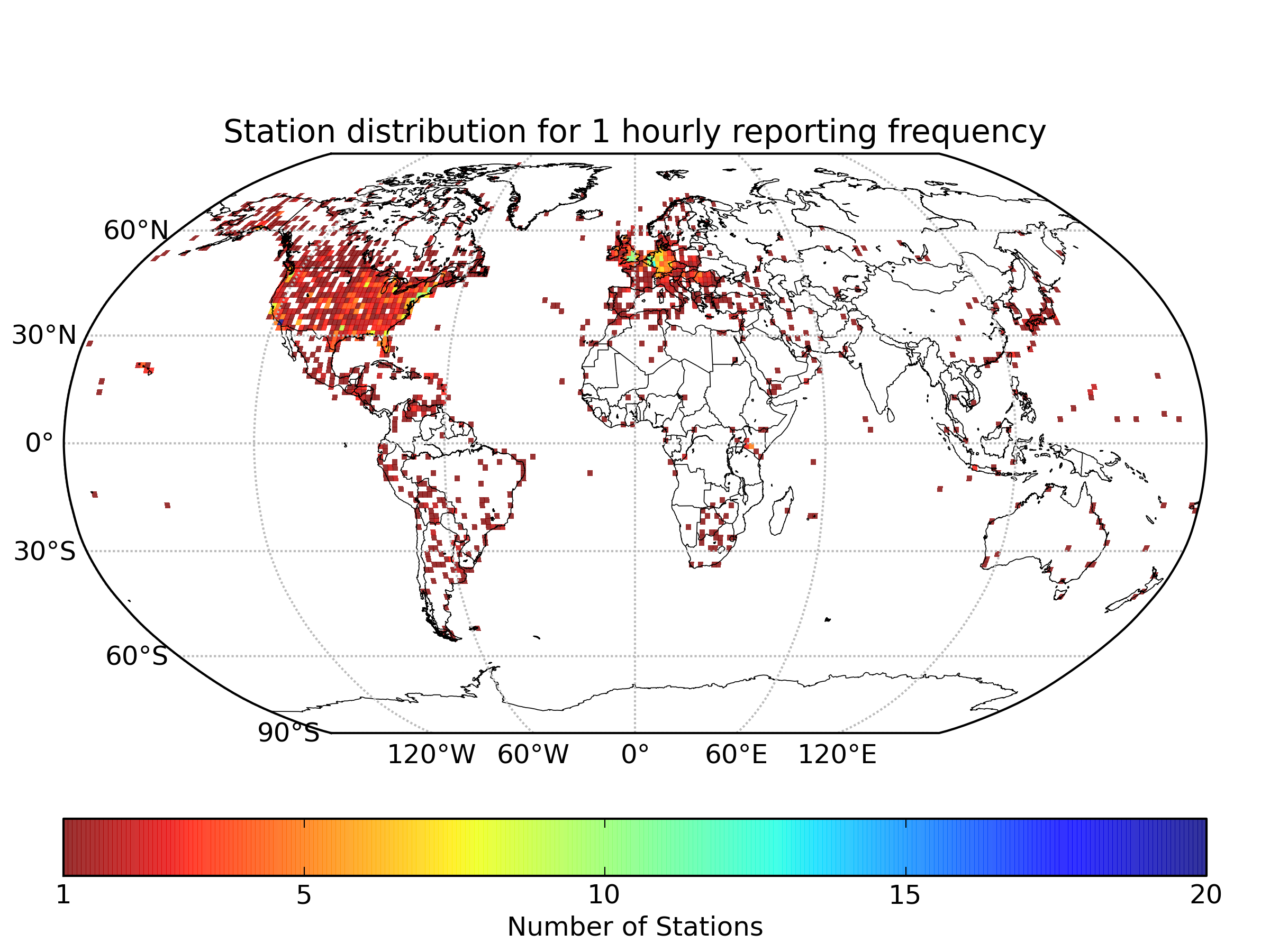}
\includegraphics[width=8.5cm]{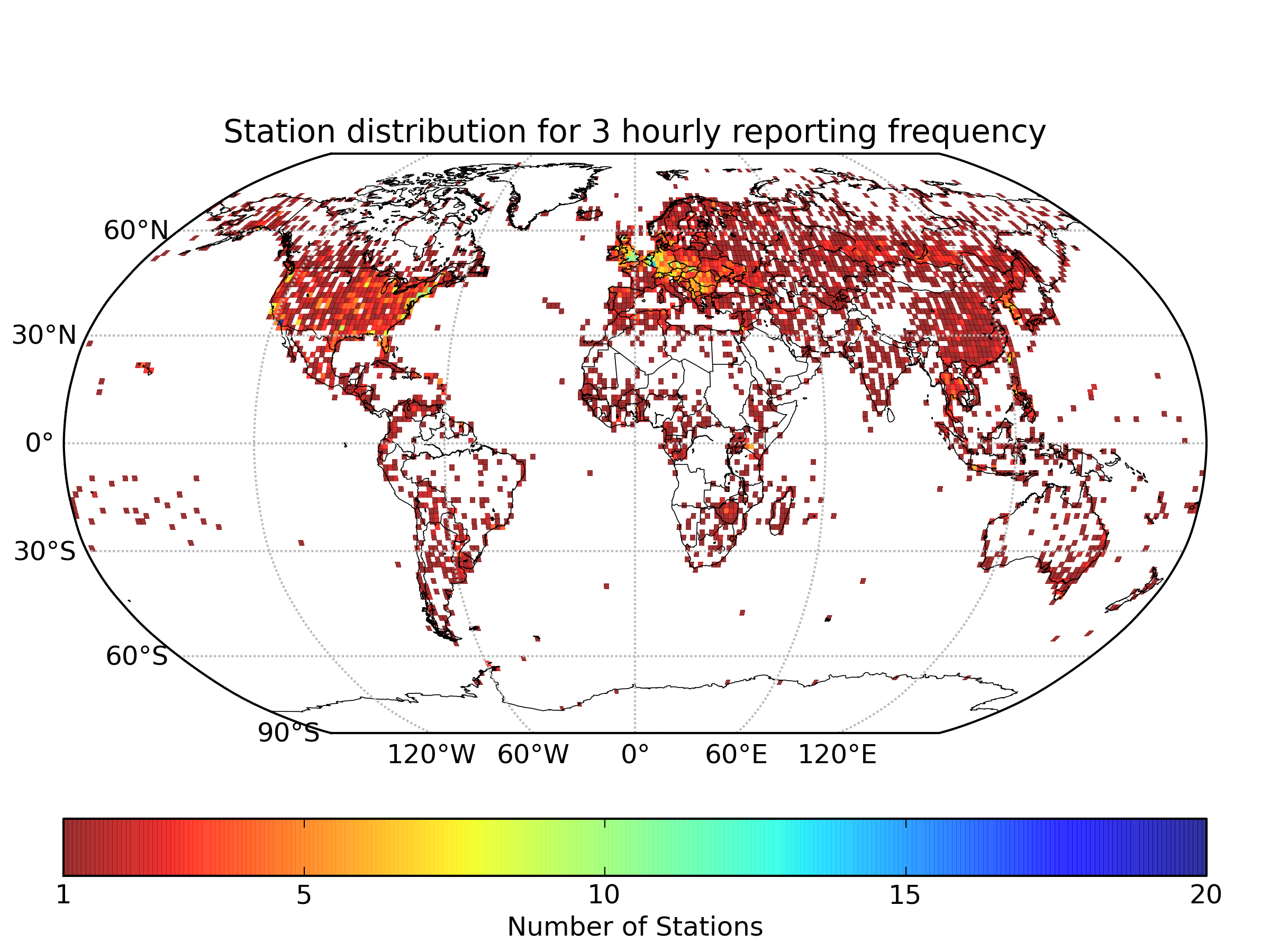}
\includegraphics[width=8.5cm]{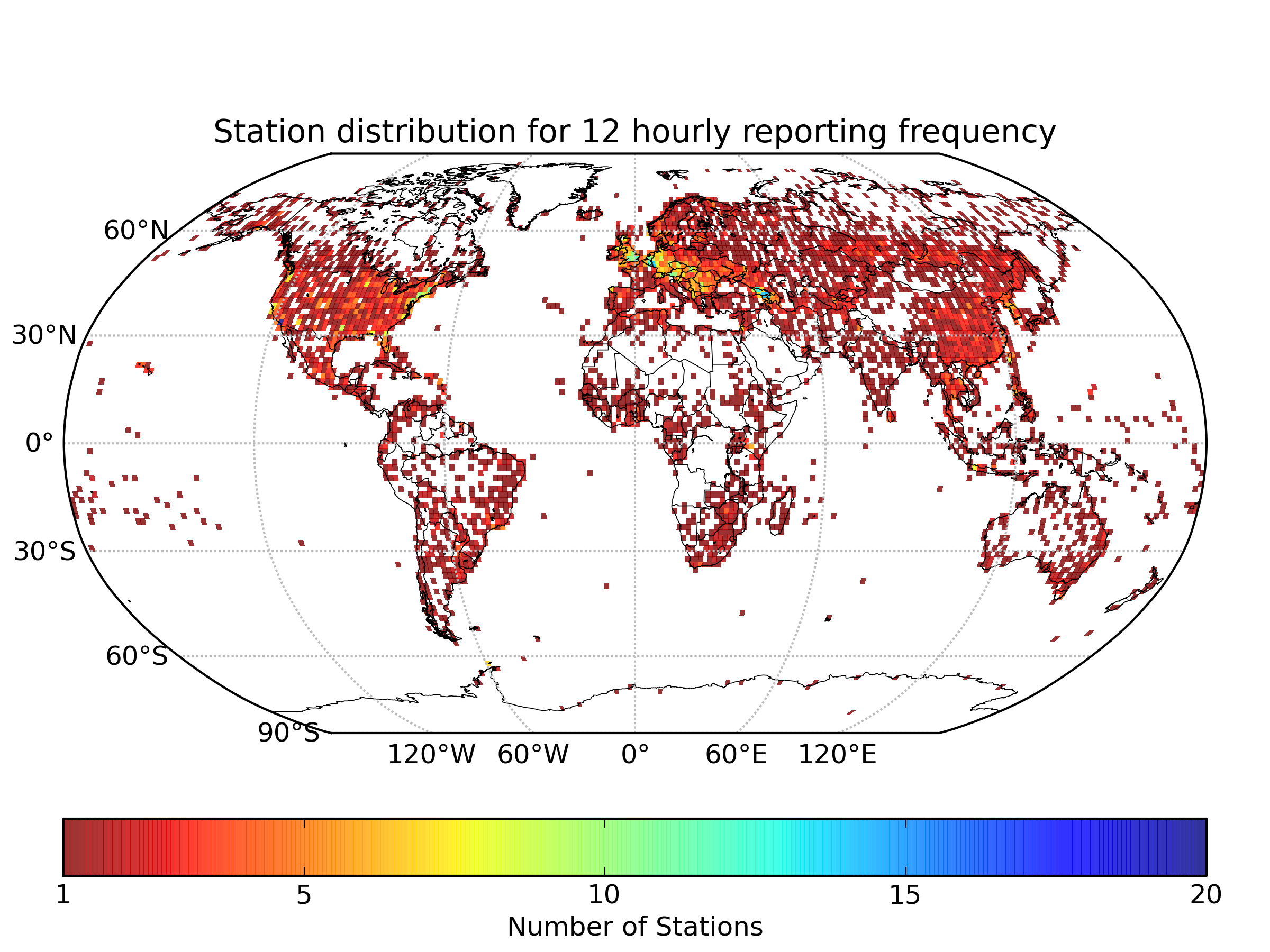}
\caption{\label{fig:2} Station distributions for different minimum
  reporting frequencies for a 1976--2005 climatology period.  For
  presentational purposes we show the number of stations within
  $1.5^\circ \times 1.5^\circ$ grid boxes. Hourly (top panel);
  3-hourly (middle panel) and 12-hourly (bottom panel).}
\end{figure}

\section{Compositing stations}\label{sec:compositing}

The ISD database archives according to the station identifier (ID)
appended to the report transmission, resulting in around 28\,000
individual station IDs. Despite efforts by the ISD dataset creators,
this causes issues for stations that have changed their reporting ID
frequently or that have reported simultaneously under multiple IDs to
different ISD source databanks (i.e.~using a WMO identifier over the
GTS and a national identifier to a local repository). Many such
station records exist in multiple independent station files within
the ISD database despite in reality being a single station record. In
some regions, e.g.~Canada and parts of Eastern Europe, WMO station ID
changes have been ubiquitous, so \mbox{compositing} is essential for record
completeness.

%t1
\begin{table}[t]
\caption{\label{table:1} Hierarchical criteria for deciding
  whether given pairs of stations in the ISD master listing were
  potentially the same station and therefore needed assessing
  further. The final value arising for a given pair of stations is the
  sum of the values for all hierarchical criteria met, e.g. a station
  pair that agrees within the elevation and latitude/longitude bounds
  but for no other criteria will have a value of 7.}
\begin{tabular}{p{120pt}l}
\tophline
&Hierarchical\\
Criteria&criteria value\\
\hhline
Reported elevation within 50\,m & \multicolumn{1}{r}{1}\\
Latitude within 0.05$^\circ$&\multicolumn{1}{r}{2}\\
Longitude within 0.05$^\circ$&\multicolumn{1}{r}{4}\\
Same country&\multicolumn{1}{r}{8}\\
WMO identifier agrees and not missing, same country&\multicolumn{1}{r}{16}\\
USAF identifier agrees in first 5 numbers and not missing&\multicolumn{1}{r}{32}\\
Station name agrees and country either the same or missing&\multicolumn{1}{r}{64}\\
METAR (Civil aviation) station call sign agrees&\multicolumn{1}{r}{128}\\
\bottomhline
\end{tabular}
\end{table}

Station location and ID information were read from the ISD station
inventory, and the potential for station matches assessed by pairwise
comparisons using a \mbox{hierarchical} \mbox{scoring} system (Table~\ref{table:1}). The
inventory is used instead of within data file location information as
the latter had been found to be substantially more questionable (Neal
Lott, personal communication, 2008). Scores are high for those elements which, if
identical, would give high confidence that the stations are the
same. For example it is highly implausible that a METAR call sign will
have been recycled between geographically distinct stations. Station
pairs that exceeded a total score of 14 are selected for further
analysis (see Table~\ref{table:1}). Therefore, a candidate pair for consideration must at
an absolute minimum be close in distance and elevation and from the
same country, or have the same ID or name. Several stations appeared
in more than one unique pairing of potential composites. These cases
were combined to form consolidated sets of potential matches. Some of
these sets comprise as many as five apparently unique station IDs in
the ISD database.

\begin{figure}[t]
\includegraphics[width=8.5cm]{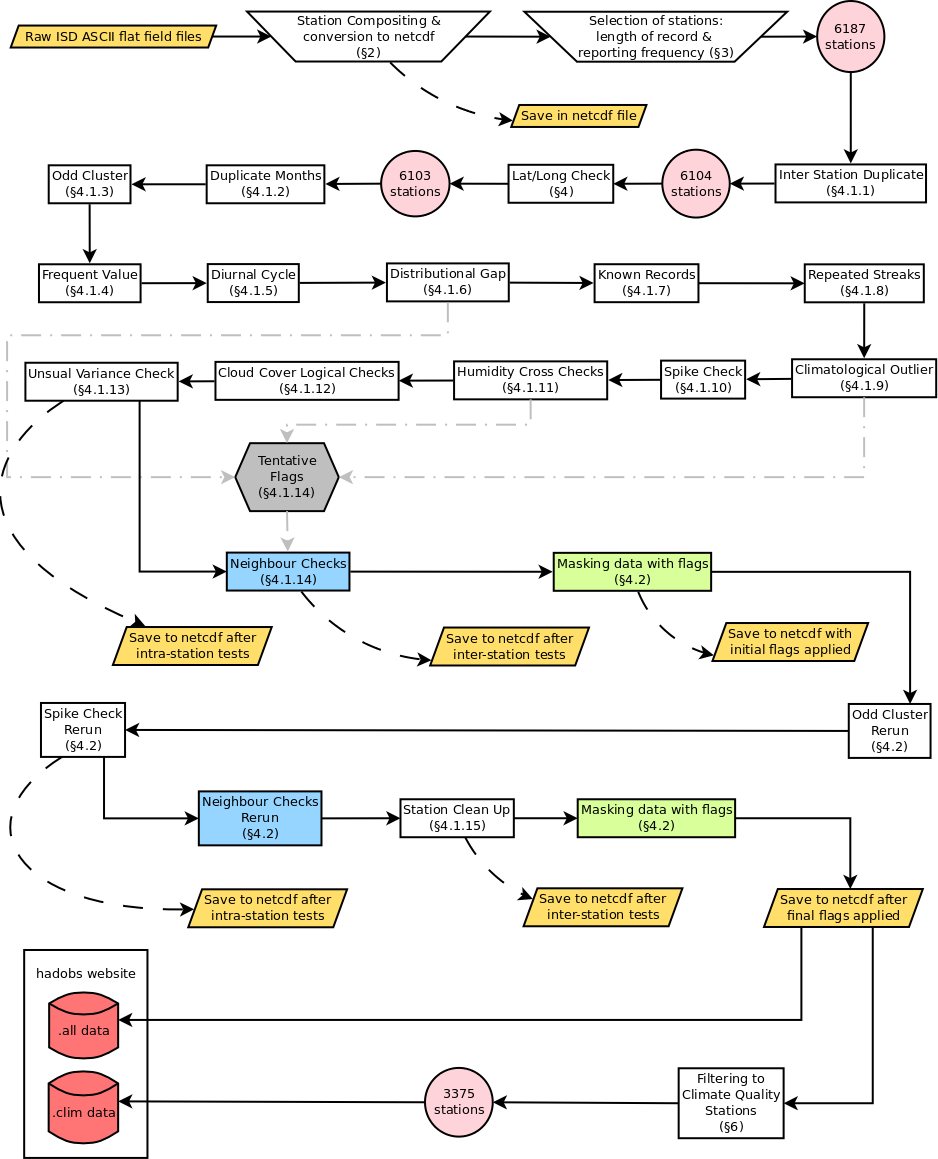}
\caption{\label{fig:3} Flow diagram of the testing procedure.  Final
  output is available on
  \url{www.metoffice.gov.uk/hadobs/hadisd}.  Other outputs (yellow
  trapezoids) are available on request.}\vspace{-3mm}
\end{figure}

\begin{figure}[t]
\includegraphics[angle=90,width=8.5cm]{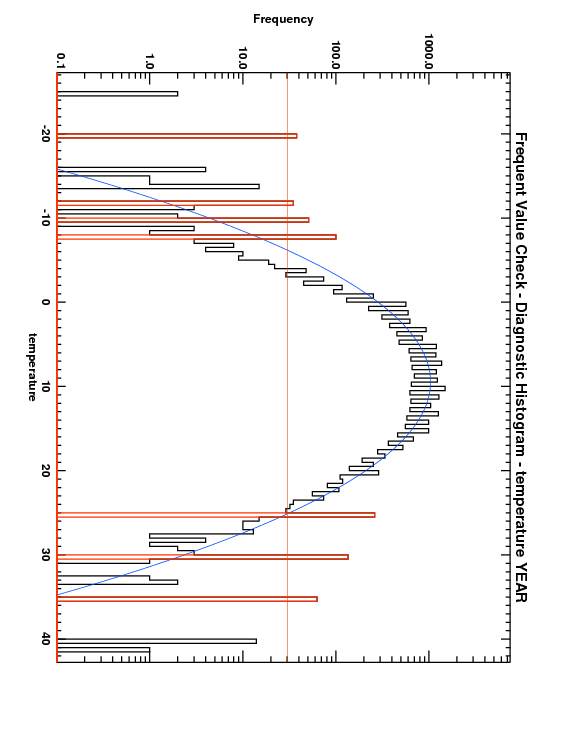}
\includegraphics[angle=90,width=8.5cm]{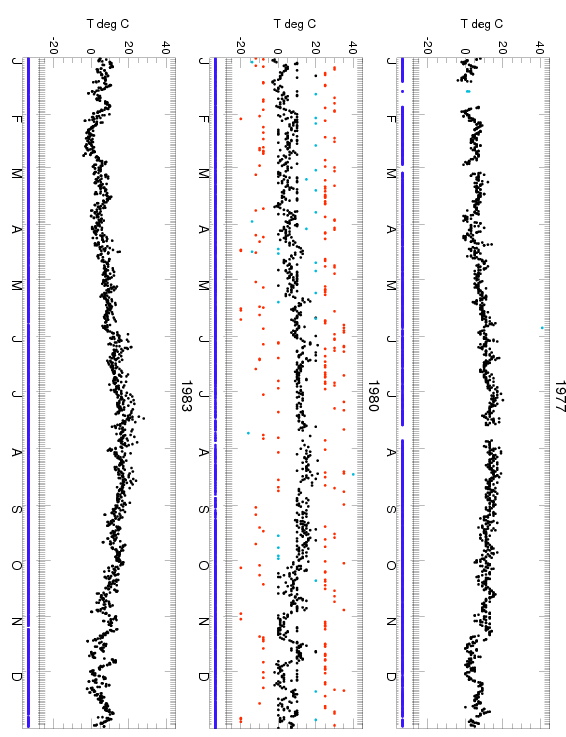}
\caption{\label{fig:4} Frequent value check (test 4) for station
  037930-99999, Anvil Green, Kent, UK (51.22$^\circ$\,N, 1.000$^\circ$\,E,
  140\,m), showing temperature. Top: Histogram with logarithmic y-axis
  for entire station record showing the bins which have been
  identified as being likely frequent values.  Bottom:~red points show
  values removed by this test and blue points by other tests for the
  years 1977, 1980 and 1983.  The panel below each year indicates
  which station the observations come from in the case of a composite
  (not relevant here but is relevant in other station plots so
  included in all).}
\end{figure}

For each potential station match set, in addition to the hierarchical
scoring system value (Table~\ref{table:1}), were considered graphically the
following quantities:~00:00\,UTC temperature anomalies from the ISD-lite
database (\url{http://www.ncdc.noaa.gov/oa/climate/isd/index.php}) using
anomalies relative to the mean of the entire set of candidate station
records; the ISD-lite data count by month;~and the daily distribution
of observing times. This required in-depth manual input taking roughly
a calendar month to complete resulting in 1504 likely composite sets
assigned as matches (comprising 3353 unique station IDs, Fig.~\ref{fig:1}). Of
these just over half are very obviously the same station. For example,
data ceased from one identifier simultaneously with data commencing
from the other where the data are clearly not substantially
inhomogeneous across the break;~or the different identifiers report at
different synoptic hours, but all other details are the same. Other
cases were less clear, in most cases because data overlap implied
potentially distinct stations or discontinuities yielding larger
uncertainties in assignment. Assigned sets were merged giving initial
preference to longer record segments but allowing infilling of missing
elements where records overlap from the shorter segment records to
maximise record completeness.  This matching of stations was carried
out on an earlier extraction of the ISD dataset spanning 1973 to 2007.
The final dataset is based on an extraction from the ISD of data
spanning 1973 to end-2011, and the station assignments have been carried
over with no reanalysis.

There may well be assigned composites that should be separate
stations, especially in densely sampled regions of the globe. If the
merge were being done for the raw ISD archive that constitutes the
baseline synoptic dataset held in the designated WMO World Data
Centre, then far more meticulous analysis would be required. For this
value added product a few false station merges can be tolerated and
later amended/removed if detected.  The station IDs that were combined
to form a single record are noted in the metadata of the final output
file where appropriate.  A list of the identifiers of the 943 stations
in the final dataset, which are assigned composites as well as their
component station IDs, can be found on the HadISD website.

\begin{figure}[t]
\includegraphics[width=8.5cm]{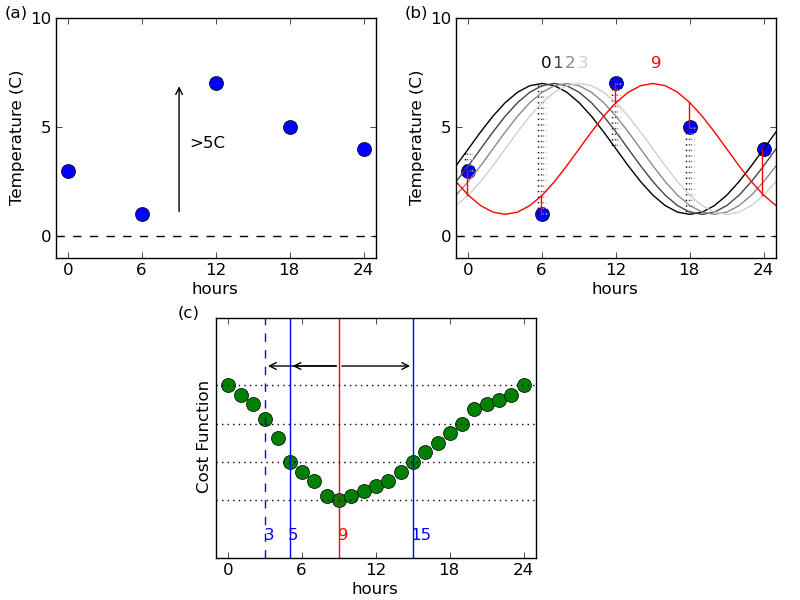}
\caption{\label{fig:5} Schematic for the diurnal cycle check.  \textbf{(a)} An
  example time series for a given day.  There are observations in more
  than 3 quartiles of the day and the diurnal range is more than
  5\,$^\circ$C so the test will run.  \textbf{(b)} A sine curve is fitted to the
  day observations.  In this schematic case, the best fit that occurs has
  a 9-h shift.  The cost function used to calculate the best fit is
  indicated by the dotted vertical lines. \textbf{(c)} The cost function
  distribution for each of the possible 24 offsets of the sine curve
  for this day.  The terciles of the distribution are shown by
  horizontal black dotted lines.  Where the cost function values enter
  the second tercile determines the uncertainty (vertical blue
  lines).  The larger of the two differences (in this case 9 to 15\,=\,6\,h) is chosen as the uncertainty.  So if the
  climatological value is between 3 and 15\,h, then this day does
  not have an anomalous diurnal cycle phase.}
\end{figure}

\begin{figure}[t]
\includegraphics[angle=90,width=8.5cm]{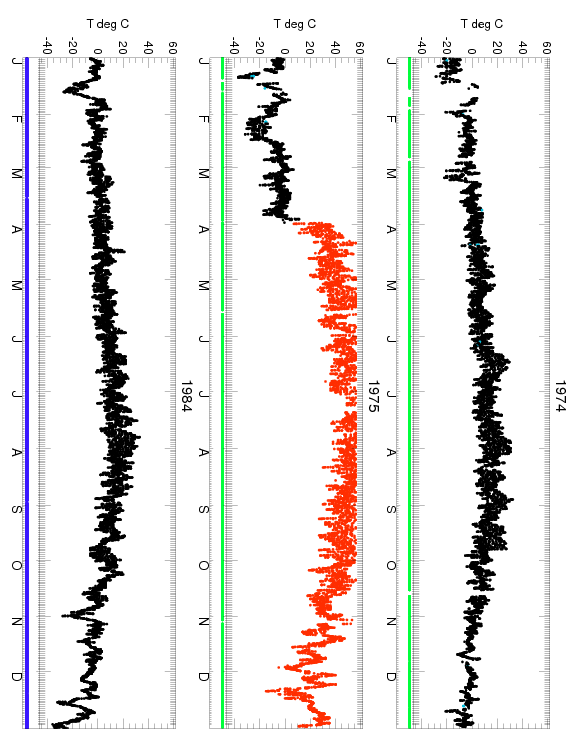}
\caption{\label{fig:6} Distributional gap check (test 6) example for
  composite station 714740-99999, Clinton, BC, Canada (51.15$^\circ$\,N,
  121.50$^\circ$\,W, 1057\,m), showing
  temperature for the years 1974, 1975 and 1984.  Red points show
  values removed by this test and blue points by other tests (none for the years shown).  The
panel below each year shows whether the data in the composited
  station come from the named station (blue) or a matched station
  (green). There is no change in source station within 1975, and so
  the compositing has not caused the clear offset observed therein,
  but the source station has changed for 1984 compared to the other
  two years.}
\end{figure}

\section{Selection and retrieval of an initial set of stations}\label{sec:selection}

The ISD consists of a large number of stations, some of which have
reported only rarely.  Of the 30\,000 stations, about 2/3 have
observations for 30\,yr or fewer and several thousand have small
total file sizes, corresponding to few observations.  However, almost
2000 stations have long records extending 60 or more years between
1901 and end-2011. Most of these have large total file sizes
indicating quasi-continuous records, rather than only a few
observations per year.  To simplify selection, only stations that
may plausibly have records suitable for climate applications were
considered, using two key requirements: length of record and reporting
frequency. The latter is important for characterisation of extremes,
as too infrequent observing will greatly reduce the potential to
capture both truly extreme events and the diurnal cycle
characteristics. A degree of pre-screening was therefore deemed
necessary prior to application of QC tests to winnow out those records
that would be grossly inappropriate for climate studies.

To maximise spatial coverage, network distributions for four
climatology periods (1976--2005, 1981--2000, 1986--2005 and 1991--2000)
and four different average time steps between consecutive reports
(hourly, 3-hourly, 6-hourly, 12-hourly) were compared. For a station
to qualify for a climatology period, at least half of the years within
the climatology period must have a corresponding data file regardless
of its size. No attempt was made at this very initial screening stage
to ensure these are well distributed within the climatological
period. To assign the reporting frequency, (up to) the first 250
observations of each annual file were used to work out the average
interval between consecutive observations. With hourly frequency,
stipulation coverage collapses to essentially NW Europe and North America
(Fig.~\ref{fig:2}). Three-hourly frequency yields a much more globally
\mbox{complete} \mbox{distribution}. There is little additional coverage or station
density derived by further coarsening to 6- (not shown) or 12-hourly
except in parts of Australia, South America and the Pacific. Sensitivity
to choice of climatology period is much smaller (not shown), so a
1976--2005 climatology period and a 3-hourly reporting frequency were
chosen as a minimum requirement.  This selection resulted in 6187
stations selected for further analysis.

%t2
\begin{table}[t]
\caption{\label{table:2} Variables extracted from the ISD database and
  converted to netCDF for subsequent potential analysis. The second
  column indicates whether the value is an instantaneous measure or a
  time-averaged quantity. The third column shows the subset that we
  quality controlled and the fourth column the set included within the
  final files which includes some non-quality controlled variables.}
\scalebox{.87}[.87]{\begin{tabular}{llll}
\tophline

&Instantaneous (I)&Subse-&Output \\
& or past period (P)&quent &in final\\

Variable                      & measurement & QC &  dataset\\
\hhline
Temperature                   &\multicolumn{1}{c}{I}& \multicolumn{1}{c}{Y}& \multicolumn{1}{c}{Y}\\
Dewpoint                      &\multicolumn{1}{c}{I}& \multicolumn{1}{c}{Y}& \multicolumn{1}{c}{Y}\\
SLP                           &\multicolumn{1}{c}{I}& \multicolumn{1}{c}{Y}& \multicolumn{1}{c}{Y}\\
Total cloud cover             &\multicolumn{1}{c}{I}& \multicolumn{1}{c}{Y}& \multicolumn{1}{c}{Y}\\
High cloud cover              &\multicolumn{1}{c}{I}& \multicolumn{1}{c}{Y}& \multicolumn{1}{c}{Y}\\
Medium cloud cover                  &\multicolumn{1}{c}{I}& \multicolumn{1}{c}{Y}& \multicolumn{1}{c}{Y}\\
Low cloud cover               &\multicolumn{1}{c}{I}& \multicolumn{1}{c}{Y}& \multicolumn{1}{c}{Y}\\
Cloud base                    &\multicolumn{1}{c}{I}& \multicolumn{1}{c}{N}& \multicolumn{1}{c}{Y}\\
Wind speed                    &\multicolumn{1}{c}{I}& \multicolumn{1}{c}{Y}& \multicolumn{1}{c}{Y}\\
Wind direction                      &\multicolumn{1}{c}{I}& \multicolumn{1}{c}{Y}& \multicolumn{1}{c}{Y}\\
Present significant weather         &\multicolumn{1}{c}{I}& \multicolumn{1}{c}{N}&\multicolumn{1}{c}{ N}\\
Past significant weather \#1        &\multicolumn{1}{c}{P}& \multicolumn{1}{c}{N}& \multicolumn{1}{c}{Y}\\
Past significant weather \#2        &\multicolumn{1}{c}{P}& \multicolumn{1}{c}{N}& \multicolumn{1}{c}{N}\\
Precipitation report \#1            &\multicolumn{1}{c}{P}& \multicolumn{1}{c}{N}& \multicolumn{1}{c}{Y}\\
Precipitation report \#2            &\multicolumn{1}{c}{P}& \multicolumn{1}{c}{N}& \multicolumn{1}{c}{N}\\
Precipitation report \#3            &\multicolumn{1}{c}{P}& \multicolumn{1}{c}{N}& \multicolumn{1}{c}{N}\\
Precipitation report \#4            &\multicolumn{1}{c}{P}& \multicolumn{1}{c}{N}& \multicolumn{1}{c}{N}\\
Extreme temperature report \#1            &\multicolumn{1}{c}{P}& \multicolumn{1}{c}{N}& \multicolumn{1}{c}{N}\\
Extreme temperature report \#2            &\multicolumn{1}{c}{P}& \multicolumn{1}{c}{N}& \multicolumn{1}{c}{N}\\
Sunshine duration             &\multicolumn{1}{c}{P}& \multicolumn{1}{c}{N}& \multicolumn{1}{c}{N}\\
\bottomhline
\end{tabular}}
\end{table}

ISD raw data files are (potentially) very large ASCII flat files --
one per station per year. The stations data were converted to hourly
resolution netCDF files for a subset of the variables including both
WMO-designated mandatory and optional reporting parameters. Details of
all variables retrieved and those considered further in the current
quality control suite are given in Table~\ref{table:2}.  There are some stations
which for part of the analysed period report at sub-hourly frequencies.  As
both temperature and dewpoint temperature are required to be measured
simultaneously for any study on humidity to be reliably carried out,
reports that have both temperature and dewpoint temperature
observations are favoured (under the assumption that the readings were
taken at close proximity in space and time) over those reports that
have one or the other (but not both), even
if the reports with both observations are further from the full hour.  In cases where observations
only have temperature or dewpoint temperature (and never both), then those with
temperature are favoured, even if these are further from the full
hour (00\,min).  All variables in a single HadISD
hourly time step always derive from a single ISD time step, with no blending
between the various within-hour reports.  However the
HadISD times are always converted to the nearest whole hour. To
minimise data storage the time axis is collapsed in the netCDF files
so that only time steps with observations are retained.

\begin{figure}[t]
\includegraphics[angle=90,width=8.5cm]{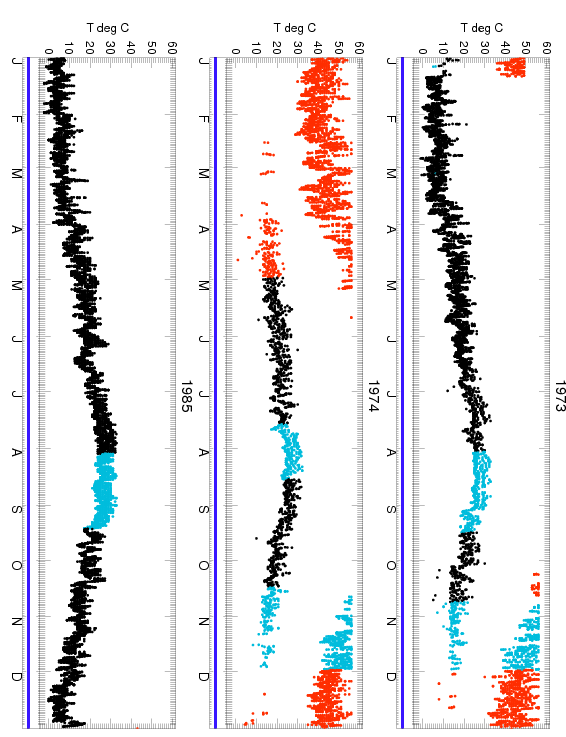}
\includegraphics[angle=90,width=8.5cm]{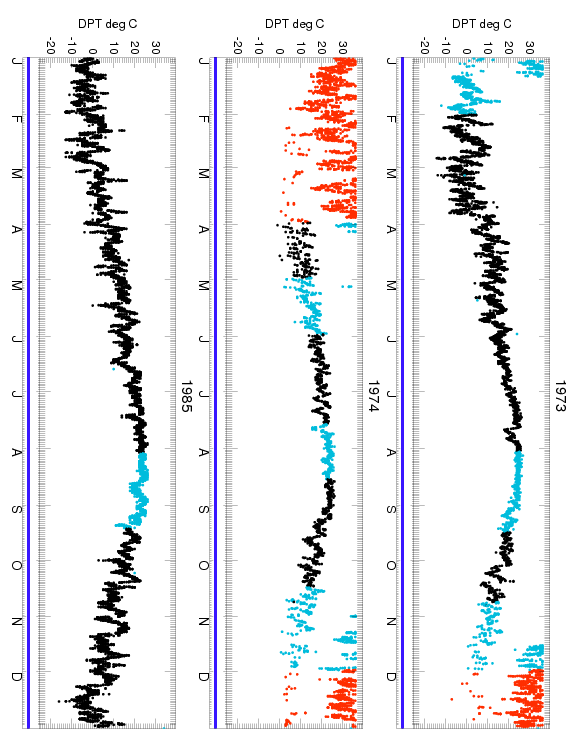}
\caption{\label{fig:7} Distributional gap check (test
  6) example when comparing all of a given calendar month in the
  dataset for composite station 476960-43323, Yokosuka, Japan (35.58$^\circ$\,N,
  139.667$^\circ$\,E, 530\,m), for
  (top) temperature and (middle) dewpoint temperature for the years
  1973, 1974 and 1985.  Red points show values removed by this test
  and blue points by other tests (in this case, mainly the diurnal
  cycle check).  The problem for this station affects
  both variables, but the tests are applied separately.  There is no
  change in source station in any of the years, and so compositing has
  not caused the bad data quality of this station. }
\end{figure}

\begin{figure}[t]
\includegraphics[angle=90,width=8.5cm]{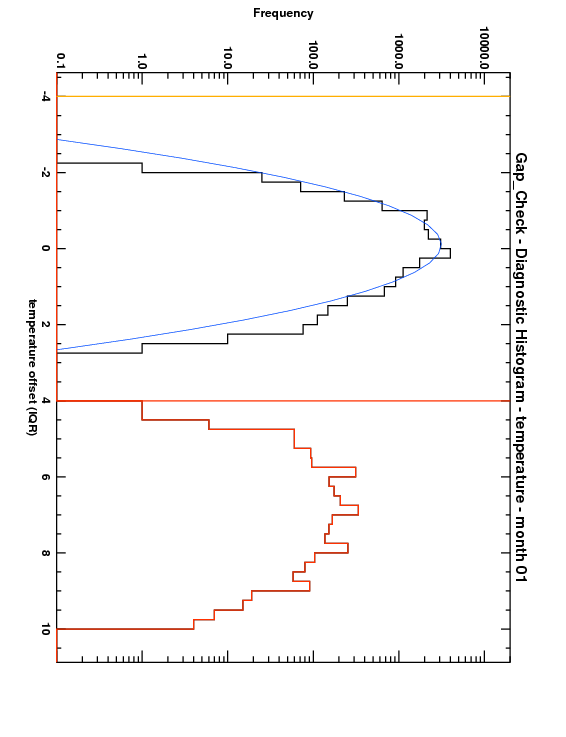}
\caption{\label{fig:7a}
  Distribution of the observations from all Januaries in the station
  record for composite station 476960-43323, Yokosuka, Japan (35.58$^\circ$\,N,
  139.667$^\circ$\,E, 530\,m).  The population highlighted in red is
  removed by the distributional gap check (test 6), as shown in Fig.~\ref{fig:7}.
  Note logarithmic y-axis.}
\end{figure}

\section{Quality control steps and analysis}\label{sec:QC}

An individual hourly station record with full temporal sampling from
1973 to 2011 could contain in excess of 340\,000 observations and there
are $>6$\,000 candidate stations. Hence, a fully automated
quality-control procedure was essential. A similar approach to that of
GHCND \citep{Durre10} was taken. Intra-station tests were
initially trained against a single (UK) case-study station series with
bad data deliberately introduced to ensure that the tests, at least to
first order, behaved as expected. Both intra- and inter-station tests
were then further designed, developed and validated based upon expert
judgment and analysis using a set of 76 stations from across the globe
(listed on the HadISD website). This set included both stations with
proportionally large data removals in early versions of the tests and
GCOS (Global Climate Observing System) Surface Network stations known
to be highly equipped and well staffed so that major problems are
unlikely. The test software suite took a number of iterations to
obtain a satisfactorily small expert judgement false positive rate
(type I error rate) and, on subjective assessment, a clean dataset for
these stations. In addition, geographical maps of detection rates were
viewed for each test and in total to ensure that rejection rates did
not appear to have a real physical basis for any given test or
variable.  Deeper validation on UK stations (IDs beginning 03) was
carried out using the well-documented 2003 heat wave and storms of
1987 and 1990.  This resulted in a further round of refining,
resulting in the tests as presented below.

Wherever distributional assumptions were made, an indicator that is
robust to outliers was required. Pervasive data issues can lead to an
unduly large standard deviation ($\sigma$) being calculated which results in
the tests being too conservative. So, the inter-quartile range (IQR)
or the median absolute deviation (MAD) was used instead; these sample
solely the (presumably reasonable) core portion of the
distribution. The IQR samples 50 per cent of the population, whereas
$\pm 1$$\sigma$ encapsulates 68 per cent of the population for a truly normal
distribution. One IQR is 1.35$\sigma$, and one MAD is 0.67$\sigma$ if the
underlying data are truly normally distributed.

\begin{figure}[t]
\includegraphics[angle=90,width=8.5cm]{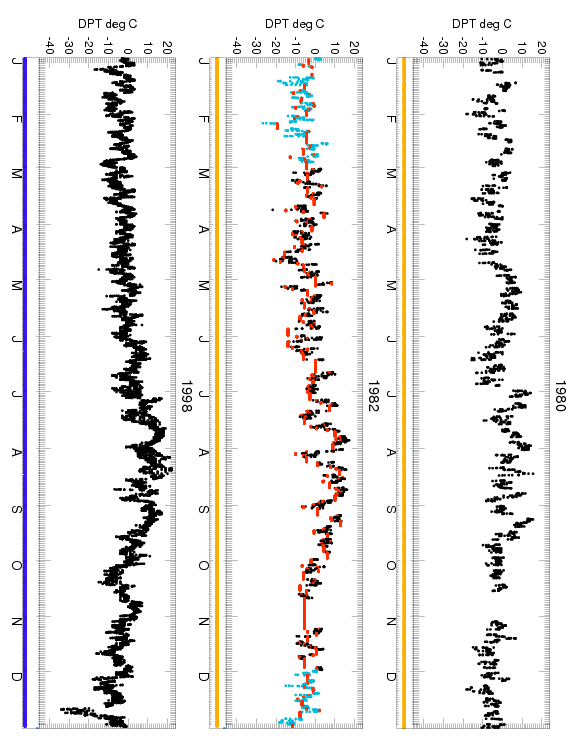}
\caption{\label{fig:8} Repeated streaks/unusual streak frequency check
  (test 8) example for composite station 724797-23176 (Milford, UT, USA;
  38.44$^\circ$\,N, 112.038$^\circ$\,W, 1534\,m), for dewpoint temperature in 1982, illustrating frequent
  short streaks.  Red points show values removed by this test and blue
  points by other tests.  The panel below each year shows whether the
  data in the composited station come from the named station (blue) or
  a matched station (orange).  There is no change in source station in
  1982, and so the compositing has not caused the streaks observed in
  1982, but a different station is used in 1998 compared to the other
  two years.}
\end{figure}

\begin{figure}[t]
\includegraphics[angle=90,width=8.5cm]{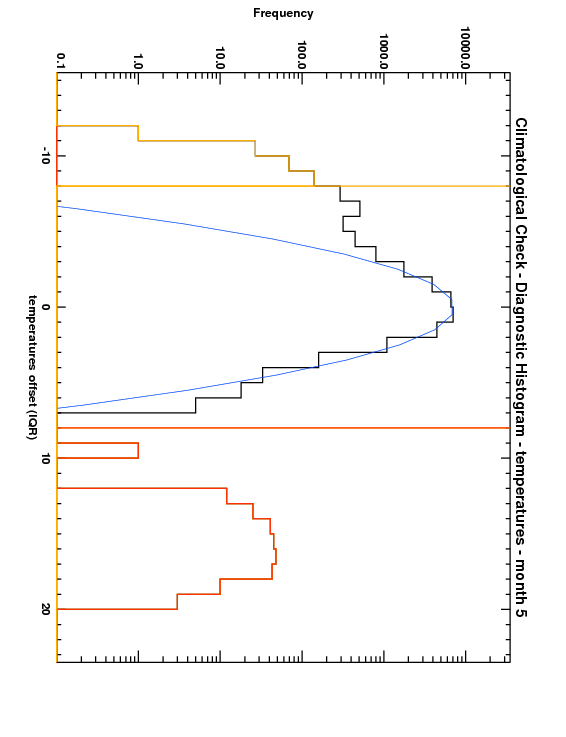}
\caption{\label{fig:9} Climatological outlier check (test 9) for
  040180-16201 (Keflavik, Iceland, 63.97$^\circ$\,N,
  22.6$^\circ$\,W, 50\,m) for temperature showing the distribution for
  May. Note logarithmic y-axis.  The threshold values are shown by the
  vertical lines.  The right-hand side shows the flagged values which
  occur further from the centre of the distribution than the gap and
  the threshold value.  The left-hand side shows observations which
  have been tentatively flagged, as they are only further from the
  centre of the distribution than the threshold value.  It is
  therefore not clear if the large tail is real or an artefact.}
\end{figure}

The \citet{Durre10} method applies tests in a deliberate order,
removing bad data progressively. Here, a slightly different approach
is taken including a multi-level flagging system. All bad data have
associated flags identifying the tests that they failed. Some tests
result in instantaneous data removal (latitude-longitude and station
duplicate checks), whereas most just flag the data. Flagged, but
retained, data are not used for any further derivations of test
thresholds. However, all retained data undergo each test such that an
individual observation may receive multiple flags. Furthermore, some
of the tests outlined in the next section set tentative flags. These
values can be reinstated using comparisons with neighbouring stations
in a later test, which reduces the chances of removing true local or
regional extremes. The tests are conducted in a specified order such
that large chunks of bad data are removed from the test threshold
derivations first and so the tests become progressively more
sensitive. After an initial latitude-longitude check (which removed
one station) and a duplicate station check, intra-station tests are
applied to the station in isolation, followed by inter-station
neighbour comparisons. A subset of the intra-station tests is then
re-run, followed by the inter-station checks again and then a final
clean-up (Fig.~\ref{fig:3}).

\subsection{QC tests}\label{sec:QC:tests}

\subsubsection{Test 1: inter-station duplicate check}\label{sec:QC:tests:1}

It is possible that two unique station identifiers actually contain
identical data. This may be simple data management error or an
artefact of dummy station files intended for temporary data
storage. To detect these, each station's temperature time series is
compared iteratively with that of every other station. To account for
reporting time ($t$) issues, the series are offset by 1\,h steps
between $t-11$ and $t+11$\,h. Series with $>1000$ coincident non-missing
data points, of which over 25 per cent are flagged as exact
duplicates, are listed for further consideration. This
computer-intensive check resulted in 280 stations being put forward
for manual scrutiny.

\begin{figure}[t]
\includegraphics[width=8.5cm]{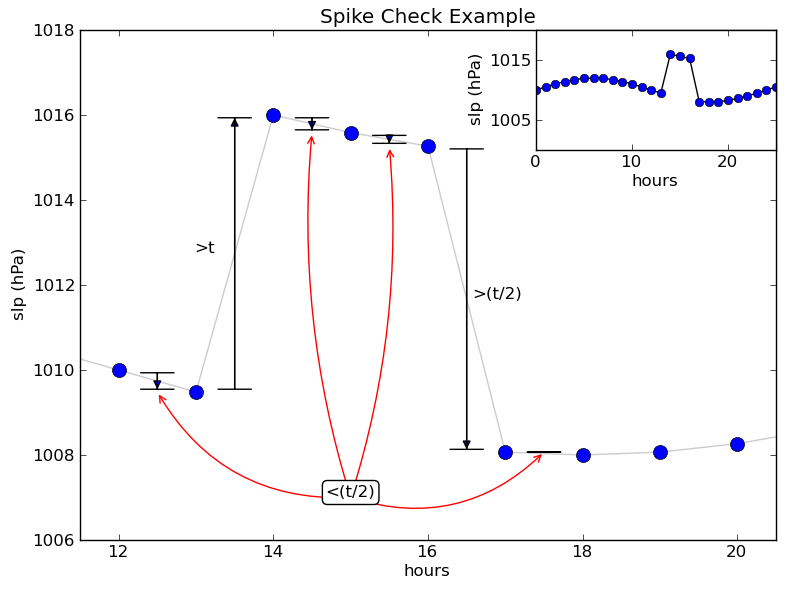}
\caption{\label{fig:10} Spike check (test 10) schematic, showing the
  requirements on the first differences inside and outside of a
  multi-point spike.  The inset shows the spike of three observations
  clearly above the rest of the time series.  The first difference
  value leading into the spike has to be greater than the threshold
  value, $t$, and the first difference value coming out of the spike has
  to be of the opposite direction and at least half the threshold
  value ($t/2$).  The differences outside and inside the spike (as
  pointed to by the red arrows) have to be less than half the
  threshold value.}
\end{figure}

All duplicate pairs and groups were then manually assessed using the
match statistics, reporting frequencies, separation distance and time
series of the stations involved. If a station pair had exact matches
on $\geq 70$ per cent of potential occasions, then the shortest station of
the pair was removed.  This results in a further loss of stations. As
this test is searching for duplicates after the merging of
composite stations (Sect.~2), any stations found by this test did
not previously meet the requirements for stations to be merged,
but still have significant periods where the observations are duplicated.  Therefore
the removal of data is the safest course of action. Stations that
appeared in the potential duplicates list twice or more were also
removed. A further subjective decision was taken to remove any
stations having a very patchy or obscure time series, for example with
very high variance. This set of checks removed a total of 83 stations
(Fig.~\ref{fig:1}), leaving 6103 to go forward into the rest of the QC
procedure.

\subsubsection{Test 2: duplicate months check}\label{sec:QC:tests:2}

Given day-to-day weather, an exact match of synoptic data for a month
with any other month in that station is highly unlikely. This test
checks for exact replicas of whole months of temperature data where at
least 20 observations are present. Each month is pattern-matched for
data presence with all other months, and any months with exact
duplicates for each matched value are flagged. As it cannot be known a
priori which month is correct, both are flagged. Although the test was
successful at detecting deliberately engineered duplication in a case
study station, no occurrences of such errors were found within the real
data. The test was retained for completeness and also because such an
error may occur in future updates of HadISD.

\begin{figure}[t]
\includegraphics[angle=90,width=8.5cm]{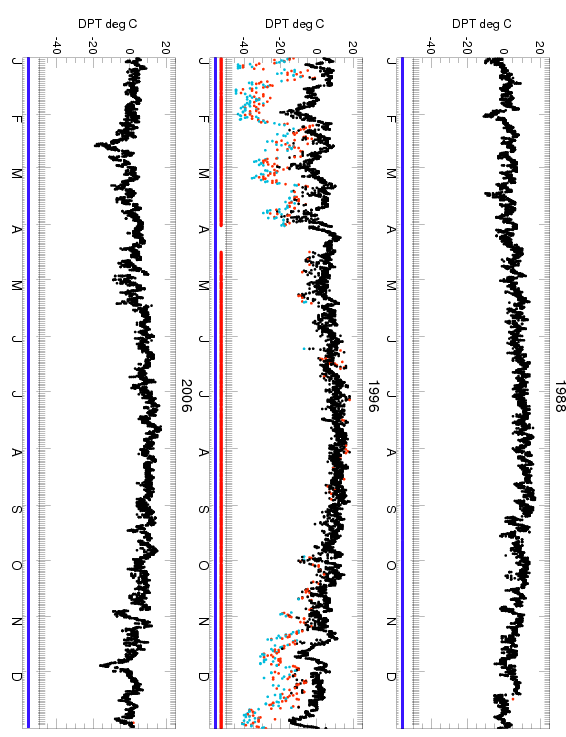}
\caption{\label{fig:11} Spike check (test 10) for composite station 718936-99999
  (49.95$^\circ$\,N, 125.267$^\circ$\,W, 106\,m, Campbell River, BC,
  Canada), for dewpoint temperature showing the removal
  of a ghost station.  Red points show values removed by this test and
  blue points by other tests. The panel below each year shows whether
  the data in the composited station come from the named station
  (blue) or a matched station (red). In 1988 and 2006 a single station
  is used for the data, but in 1996 there is clearly a blend between
  two stations (718936-99999 and 712050-99999).  In this case the
  compositing has caused the ghosting; however, both these
  stations are labelled in the ISD history file as
  Campbell River, with identical latitudes and longitudes.  An
  earlier period of merger between these two stations did not lead to any
  ghosting effects.}
\end{figure}

\begin{figure}[t]
\includegraphics[angle=90,width=8.5cm]{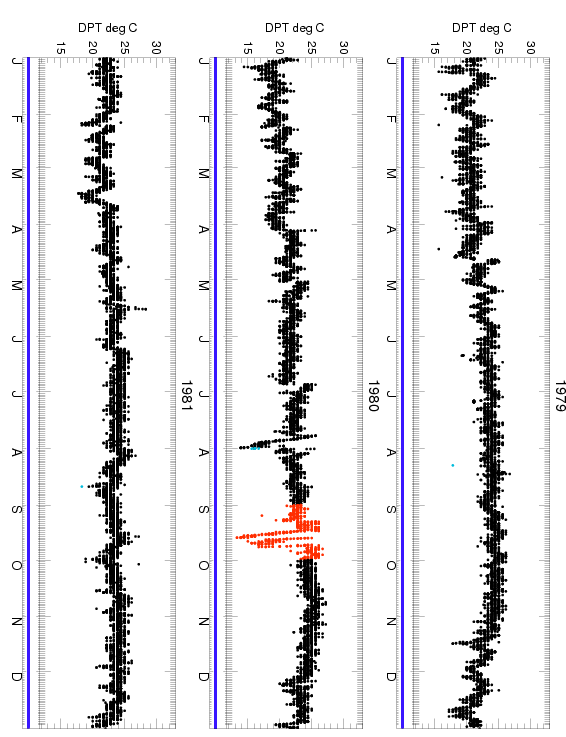}
\includegraphics[angle=90,width=8.5cm]{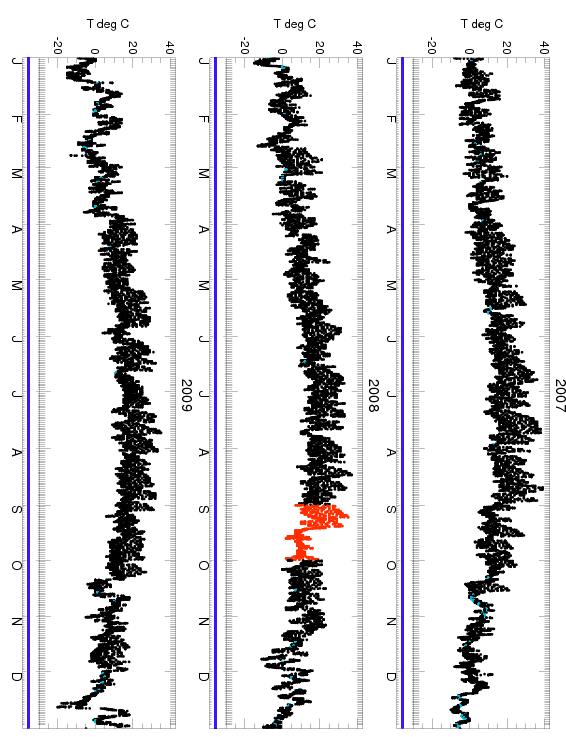}
\caption{\label{fig:12} Unusual variance check (test 13) for
  (top) 912180-99999 (13.57$^\circ$\,N, 144.917$^\circ$\,E, 162\,m, Andersen
  Air Force Base, Guam) for dewpoint temperature and (bottom)
  133530-99999 (43.82$^\circ$\,N, 18.33$^\circ$\,E, 511\,m, Sarajevo,
  Bosnia-Herzegovina) for temperature. Red points show values removed
  by this test and blue points by other tests (none for the years and variables
shown).}
\end{figure}

\subsubsection{Test 3: odd cluster check}\label{sec:QC:tests:3}

A number of time series exhibit isolated clusters of data. An
instrument that reports sporadically is of questionable scientific
value. Furthermore, with little or no surrounding data it is much more
difficult to determine whether individual observations are
valid. Hence, any short clusters of up to 6\,h within a 24\,h
period separated by 48\,h or longer from all other data are
flagged. This applies to temperature, dewpoint temperature and
sea-level pressure elements individually. These flags can be undone if
the neighbouring stations have concurrent, unflagged observations
whose range encompasses the observations in question (see Sect.~\ref{sec:QC:tests:14}).

\subsubsection{Test 4: frequent value check}\label{sec:QC:tests:4}

The problem of frequent values found in \citet{Durre10} also
extends to synoptic data. Some stations contain far more observations
of a given value than would be reasonably expected. This could be the
use of zero to signify missing data, or the occurrence of some other
local data-issue identifier\footnote{A ``local data-issue identifier'' is where
a physically valid but locally implausible value is used to mark a problem with a
particular data point.  On subsequent ingestion into the ISD, this value has been
interpreted as a real measurement rather than a flag.} that has been mistakenly ingested into
the database as a true value.  This test identifies suspect values
using the entire record and then scans for each value on a
year-by-year basis to flag only if they are a problem within that
year.

This test is also run seasonally (JF\,+\,D, MAM, JJA, SON), using a
similar approach as above.  Each set of three months is scanned over
the entire record to identify problem values (e.g.~all MAMs over the
entire record), but flags applied on an annual basis using just the
three months on their own (e.g.~each MAM individually, scanning for
values highlighted in the previous step). As indicated by JF\,+\,D, the
January and February are combined with the following December (from
the same calendar year) to create a season, rather than working with
the December from the previous calendar year.  Performing a seasonal
version, although having fewer observations to work with, is more
powerful because the seasonal shift in the distribution of the
temperatures and dewpoints can reveal previously hidden frequent
values.

For the filtered (where previously flagged observations are not
included) temperature, dewpoint and sea-level pressure data,
histograms are created with 0.5 or 1.0\,$^\circ$C or hPa increments (depending
on the reporting accuracy of the measurement) and each histogram bin
compared to the three on either side. If this bin contains more than
half of the total population of the seven bins combined and also more
than 30 observations over the station record (20 for the seasonal
scan), then the histogram bin interval is highlighted for further
investigation (Fig.~\ref{fig:4}). The minimum number limit was imposed to
avoid removing true tails of the distribution.

\begin{figure}[t]
\includegraphics[angle=90,width=8.5cm]{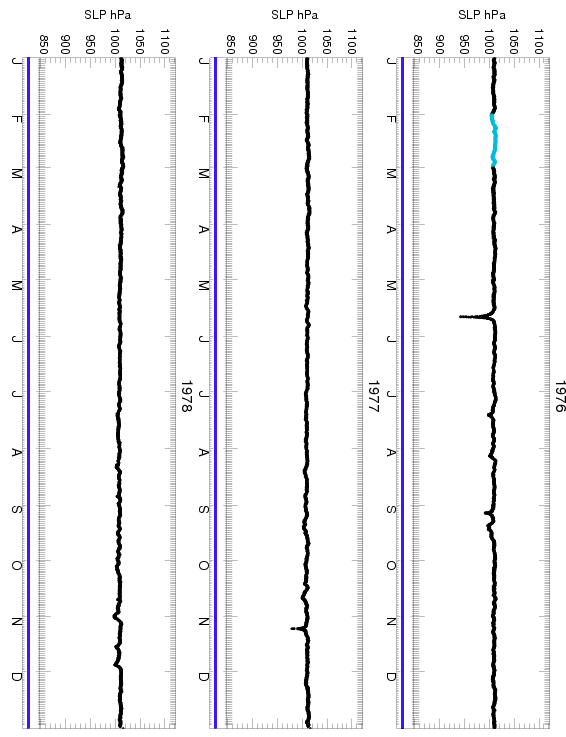}
\caption{\label{fig:13} Nearest neighbour data check (test 14) for
  912180-99999 (13.57$^\circ$\,N, 144.917$^\circ$\,E, 162\,m,
  Andersen Air Force Base, Guam) for sea-level pressure.  Red points
  show values removed by this test (none for the years
shown) and blue points by other tests.  The spikes for the
hurricanes in 1976 and 1977 are kept in the dataset.
  February 1976 is removed by the variance check -- this February has
  higher variance than expected when compared to all other Februaries
  for this station.}
\end{figure}

After this identification stage, the unfiltered distribution is
studied on a yearly basis. If the highlighted bins are prominent
(contain $>50$ per cent of the observations of all seven bins and more
than 20 observations in the year, or $90$ per cent of the observations
of all seven bins and more than 10 observations in the year) in any
year, then they are flagged (the bin sizes are reduced to 15 and 10
respectively for the seasonal scan). This two-stage process was
designed to avoid removing too many valid observations (type II
errors). However, even with this method, by flagging all values within
a bin it is likely that some real data are flagged if the values are
sufficiently close to the mean of the overall data distribution. Also,
frequent values that are pervasive for only a few years out of a
longer record and are close to the distribution peak may not be
identified with this method (type I errors). However, alternative
solutions were found to be too computationally inefficient. Station
037930-99999 (Anvil Green, Kent, UK) shows severe problems from
frequent values in the temperature data for 1980
(Fig.~\ref{fig:4}). Temperature and dewpoint flags are
synergistically applied,
i.e.~temperature flags are applied to both temperature and dewpoint
data, and vice versa.

\subsubsection{Test 5: diurnal cycle check}\label{sec:QC:tests:5}

All ISD data are archived as UTC; conversion has generally taken place
from local time at some point during recording, reporting and
archiving the data. Errors could introduce large biases into the data
for some applications that consider changes in the diurnal
characteristics. The test is only applied to stations at latitudes
below 60$^\circ$\,N/S as above these latitudes the diurnal cycle in temperature
can be weak or absent, and obvious robust geographical patterns across
political borders were apparent in the test failure rates when it was
applied in these regions.

This test is run on temperature only as this variable has the most
robust diurnal cycle, but it flags data for all variables.  Firstly, a
diurnal cycle is calculated for each day with at least four
observations spread across at least three quartiles of the day (see
Fig.~\ref{fig:5}). This is done by fitting a sine curve with amplitude equal
to half the spread of reported temperatures on that day.  The phase of
the sine curve is determined to the nearest hour by minimising a cost
function, namely the mean squared deviations of the observations from
the curve (see Fig.~\ref{fig:5}). The climatologically expected phase for a
given calendar month is that with which the largest number of
individual days phases agrees. If a day's temperature range is less
than 5\,$^\circ$C, no attempt is made to determine the diurnal cycle for that
day.

\begin{figure*}%[p]
\centering
\includegraphics[angle=90,width=13cm]{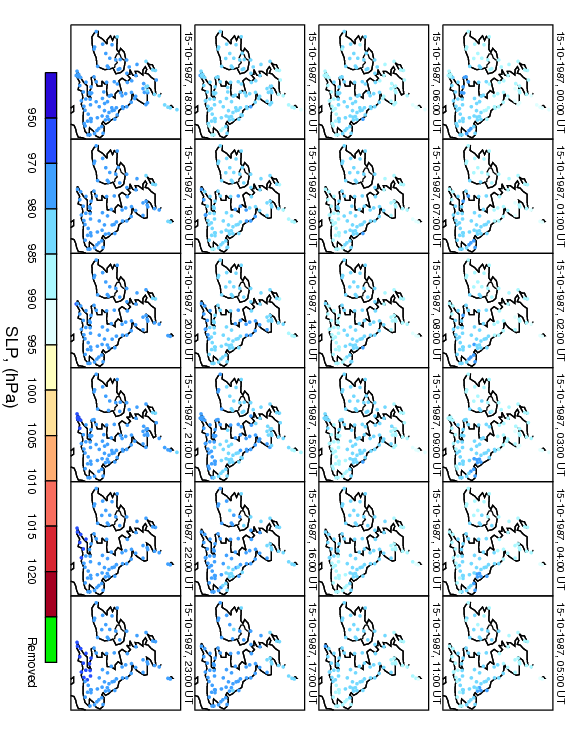}\\
\includegraphics[angle=90,width=13cm]{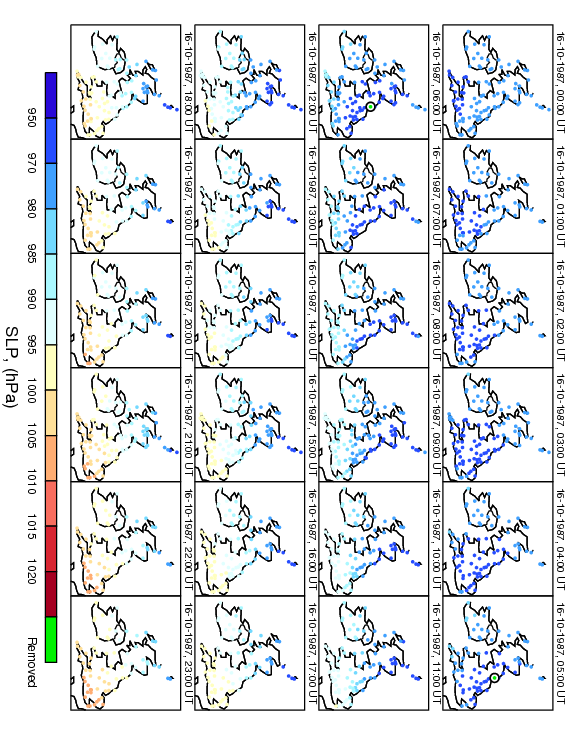}
\caption{\label{fig:14} Passage of low pressure core over the British
  Isles during the night of 15--16~October~1987.  Green points
  (highlighted by circles) are stations where the observation for that
  hour has been removed.  There are two, at 05:00 and 06:00\,UTC, on 16~October~1987 in the north-east of England.  These flagged
  observations are investigated in Fig.~\ref{fig:15}.}
\end{figure*}

\begin{figure}%[t]
\includegraphics[angle=90,width=8.5cm]{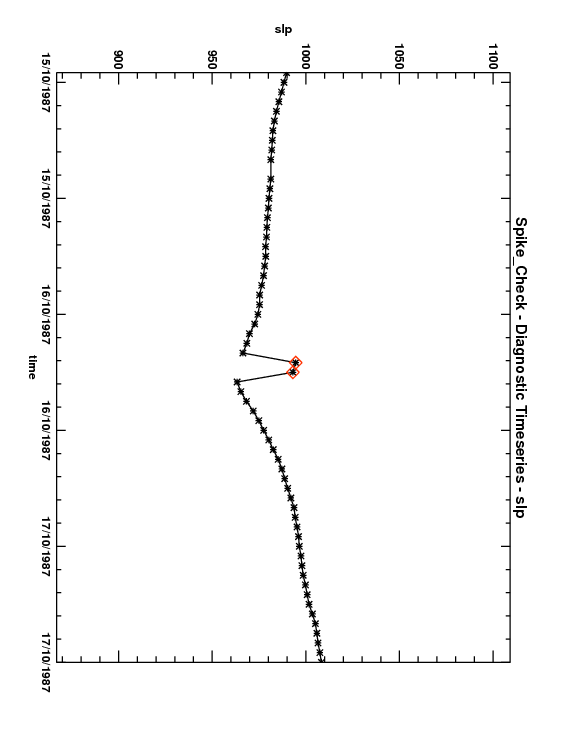}
\caption{\label{fig:15} Sea level pressure data from station
  032450-99999 (Newcastle Weather Centre, 54.967$^\circ$\,N, $-$1.617$^\circ$\,W,
  47\,m) during mid-October~1987.  The two observations
  that have triggered the spike check are clearly visible and are
  distinct from the rest of the data.  Given their values (994.6 and
  993.1\,hPa), the two flagged observations are clearly separate from
  their adjacent ones (966.4 and 963.3\,hPa). It is possible that a
  keying error in the SYNOP report led to 946 and 931 being reported,
  rather than 646 and 631.  However, we make no attempt in this
  dataset to rescue flagged values.}
\end{figure}

It is then assessed whether a given day's fitted phase matches the
expected phase within an uncertainty estimate. This uncertainty
estimate is the larger of the number of hours by which the day's phase
must be advanced or retarded for the cost function to cross into the
middle tercile of its distribution over all 24 possible phase-hours
for that day. The uncertainty is assigned as symmetric (see Fig.~\ref{fig:5}). Any periods $>30$ days where the diurnal cycle deviates from the
expected phase by more than this uncertainty, without three
consecutive good or missing days or six consecutive days consisting of
a mix of only good or missing values, are deemed dubious and the
entire period of data (including all non-temperature elements) is
flagged.

Small deviations, such as daylight saving time (DST) reporting hour
changes, are not detected by this test.  This type of problem has been
found for a number of Australian stations where during DST the local
time of observing remains constant, resulting in changes in the common
GMT reporting hours across the year\footnote{Such an error has been noted and reported back to the ISD team at NCDC.}.  Such changes in reporting
frequency and also the hours on which the reports are taken are noted
in the metadata of the netCDF file.

\begin{figure*}[p]
\centering
\includegraphics[angle=90,width=13cm]{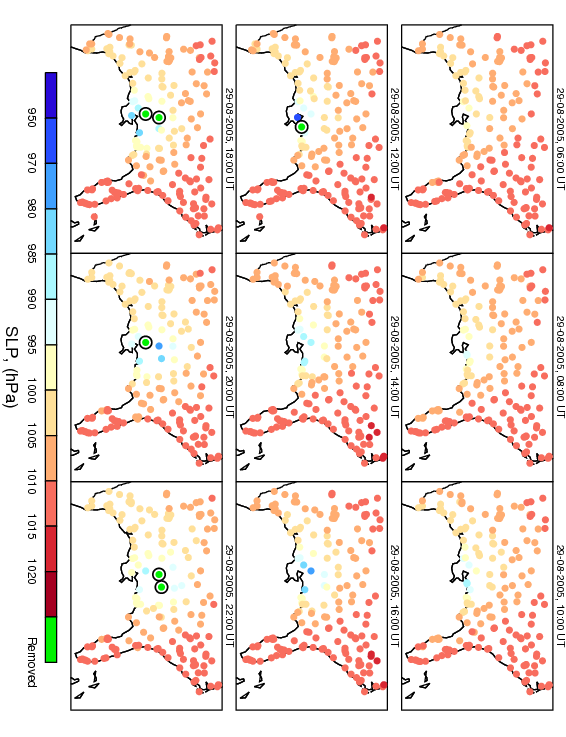}\\
\includegraphics[angle=90,width=13cm]{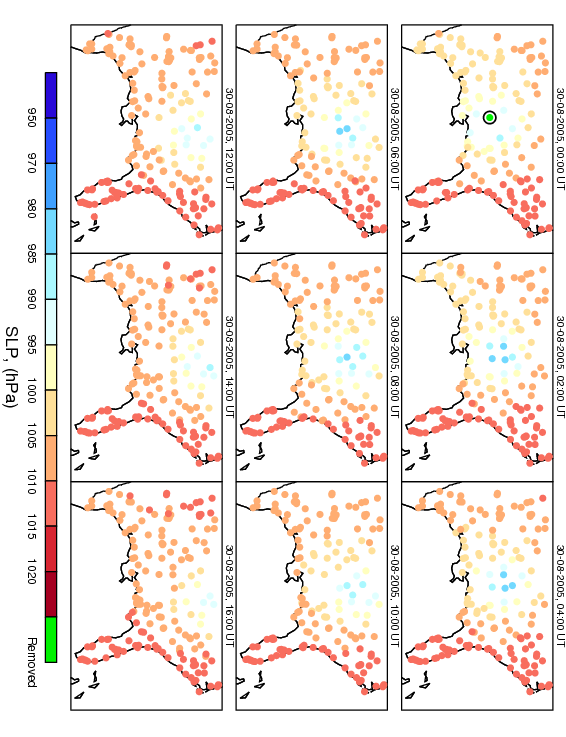}
\caption{\label{fig:16} Passage of low pressure core of Hurricane
  Katrina during its landfall in 2005.  Every second hour is
  shown. Green points are observations which have been removed, in
  this case by the neighbour outlier check (see test 14).}
\end{figure*}

\subsubsection{Test 6: distributional gap check}\label{sec:QC:tests:6}

Portions of a time series may be erroneous, perhaps originating from
station ID issues, recording or reporting errors, or instrument
malfunction. To capture these, monthly medians $M_{ij}$ are created from
the filtered data for calendar month $i$ in year $j$. All monthly medians
are converted to anomalies $A_{ij} \equiv M_{ij} - M_i$ from the calendar monthly
median $M_i$ and standardised by the calendar month inter-quartile range
IQR$_i$ (inflated to 4\,$^\circ$C or hPa for those months with very small IQR$_i$) to
account for any seasonal cycle in variance. The station's series of
standardised anomalies $S_{ij} A_{ij} / \textrm{IQR}_i$ is then ranked, and the
median, $\acute{S}$, obtained.

\begin{figure*}[t]
\centering
\includegraphics[angle=90,width=8cm]{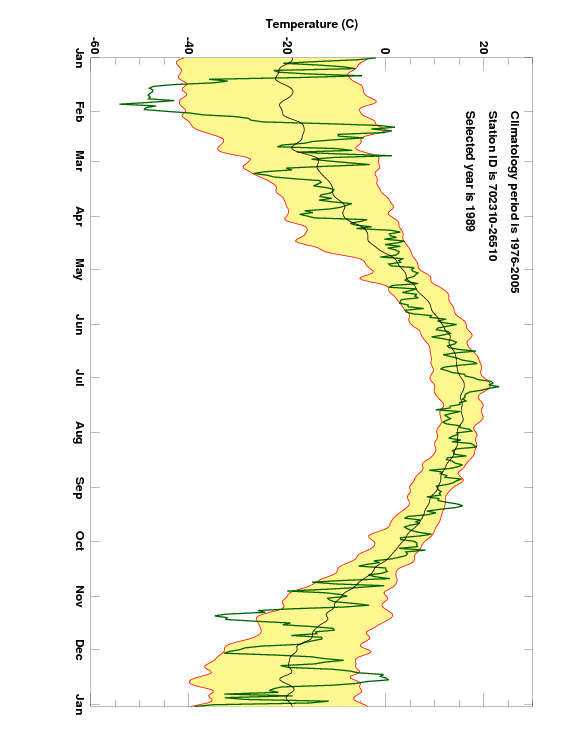}
\includegraphics[angle=90,width=8cm]{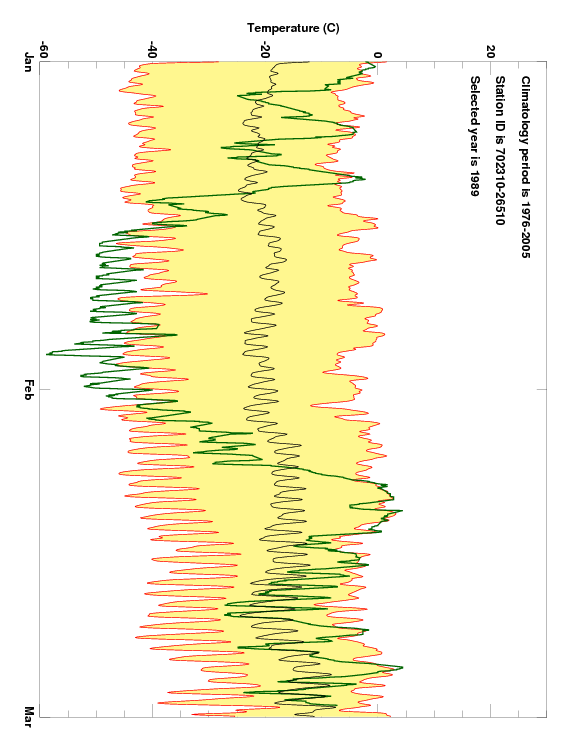}\\
\includegraphics[angle=90,width=8cm]{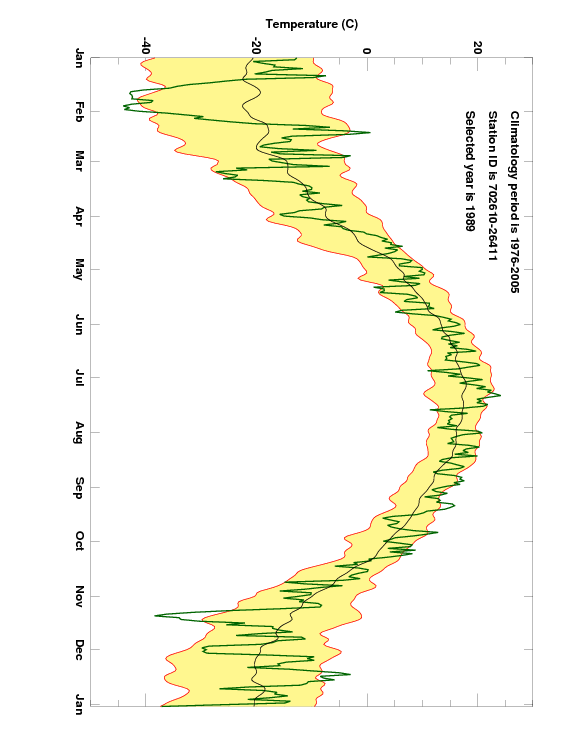}
\includegraphics[angle=90,width=8cm]{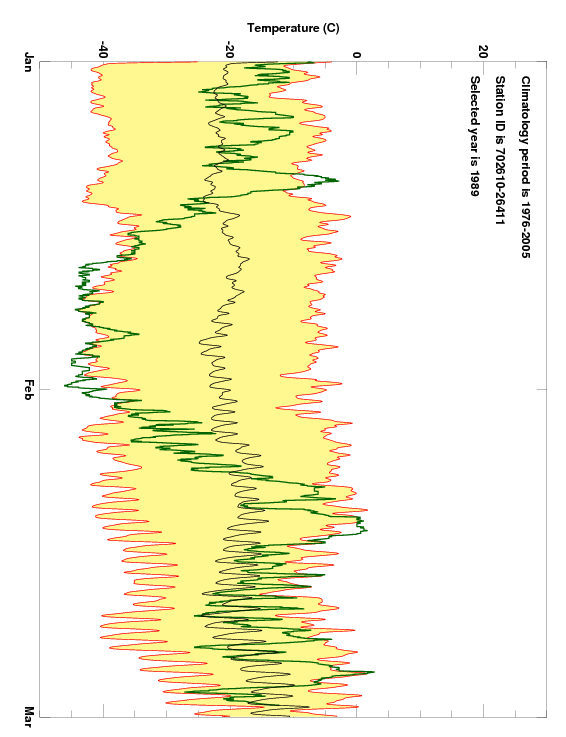}
\caption{\label{fig:17} Left: Alaskan daily mean temperature in 1989
  (green curve) shown against the climatological daily average
  temperature (black line) and the 5th and 95th percentile region, red
  curves and yellow shading.  The cold spell in late January is
  clearly visible.  Right:~similar plots, but showing the sub-daily
  resolution data for a two month period starting in January
  1989.  The climatology, 5th and 95th percentile lines have been
  smoothed using an 11-point binomial filter in all four plots. Top:
  McGrath (702310-99999, 62.95$^\circ$\,N, 155.60$^\circ$\,W, 103\,m),
  bottom: Fairbanks (702610-26411, 64.82$^\circ$\,N, 147.86$^\circ$\,W,
  138\,m). }\vspace{3mm}
\end{figure*}

Firstly, all observations in any month and year with $S_{ij}$ outside the
range $\pm5$ (in units of the IQR$_i$) from $\acute{S}$ are flagged, to remove gross
outliers.  Then, proceeding outwards from $\acute{S}$, pairs of $S_{ij}$ above and
below ($S_{iu}$, $S_{iv}$) it are compared in a step-wise fashion.  Flagging is
triggered if one anomaly $S_{iu}$ is at least twice the other $S_{iv}$ and both
are at least 1.5IQR$_i$ from $\acute{S}$.  All observations are flagged for the
months for which $S_{ij}$ exceeds $S_{iu}$ and has the same sign. This flags one
entire tail of the distribution.  This test should identify stations
that have a gap in the data distribution, which is unrealistic. Later
checks should find any issues existing in the remaining tail. Station 714740-99999
(Clinton, BC, Canada, an assigned composite) shows an example of the
effectiveness of this test at highlighting a significantly outlying
period in temperature between 1975 and 1976 (Fig.~\ref{fig:6}).

\begin{figure*}[t]
\centering
\includegraphics[angle=90,width=8cm]{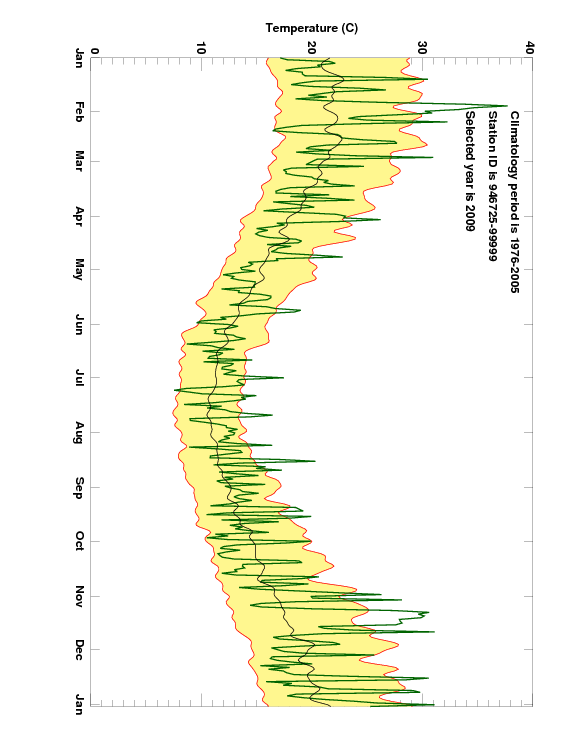}
\includegraphics[angle=90,width=8cm]{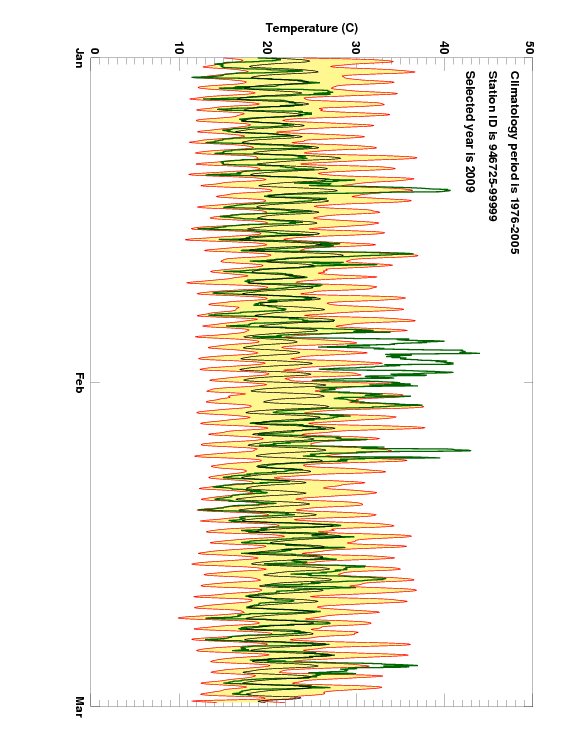}\\
\includegraphics[angle=90,width=8cm]{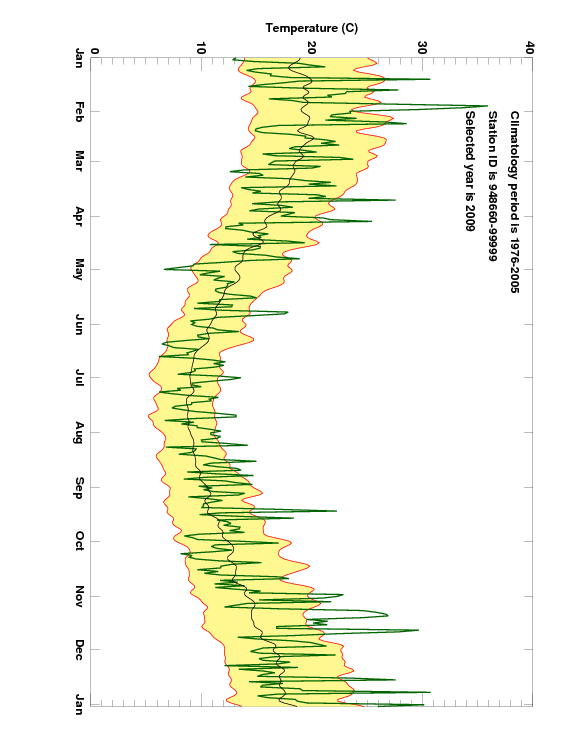}
\includegraphics[angle=90,width=8cm]{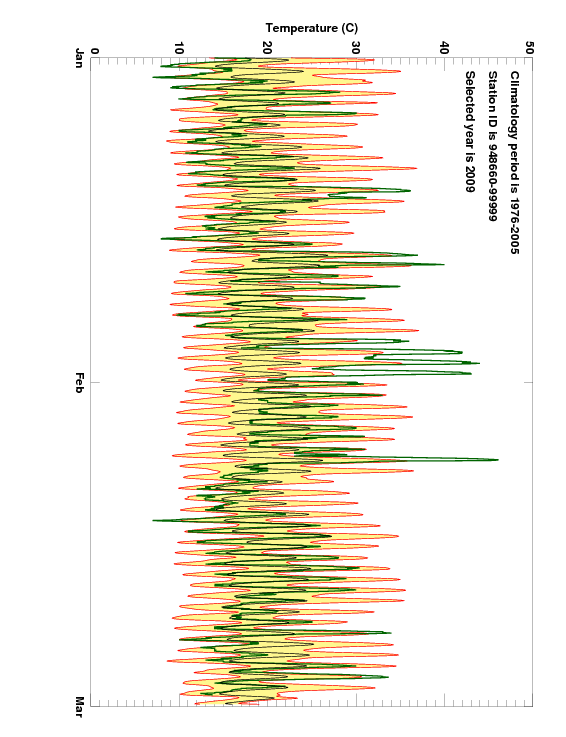}
\caption{\label{fig:18} Left: daily mean temperature in southern
  Australia in 2009 (green curve) with climatological average (black
  line) and 5th and 95th percentiles (red lines and yellow shading).
  The exceptionally high temperatures in late January/early February
  and mid-November can clearly be seen.  Right:~similar plots showing
  the full sub-daily resolution data for a two month period starting in
  January~2009.  The climatology, 5th and 95th percentile lines have
  been smoothed using an 11-point binomial filter in all four
  plots. Top: Adelaide (946725-99999, 34.93$^\circ$\,S, 138.53$^\circ$\,E,
  4\,m), bottom:~Melbourne (948660-99999, 37.67$^\circ$\,S,
  144.85$^\circ$\,E, 119\,m). }\vspace{4mm}
\end{figure*}

An extension of this test compares all the observations for a given
calendar month over all years to look for outliers or secondary
populations.  A histogram is created from all observations within a
calendar month.  To characterise the width of the distribution for
this month, a Gaussian curve is fitted.  The positions where this
expected distribution crosses the $y=0.1$ line are noted\footnote{When
  the Gaussian crosses the $y=0.1$ line, assuming a Gaussian
  distribution for the data, the expectation is that there would be
  less than 1/10th of an observation in the entire data
  series for values beyond this point for this data distribution.
  Hence we would not expect to see any observations in the data
  further from the mean if the distribution was perfectly Gaussian.
  Therefore, any observations that are significantly further from the
  mean and are separated from the rest of the observations may be
  suspect.  In Fig.~\ref{fig:7} this crossing occurs at around 2.5IQR.
  Rounding up and adding one results in a threshold of 4IQR.  There is
  a gap of greater than 2 bin widths prior to the beginning of the
  second population at 4IQR, and so the secondary population is
  flagged.}, and rounded
outwards to the next integer-plus-one to create a threshold value.
From the centre outwards, the histogram is scanned for gaps, i.e.~bins which have a value of zero.  When a gap is found, and it is large
enough (at least twice the bin width), then any bins beyond the end of
the gap, which are also beyond the threshold value, are flagged.

\begin{figure}[p]
\includegraphics[angle=90,width=8cm]{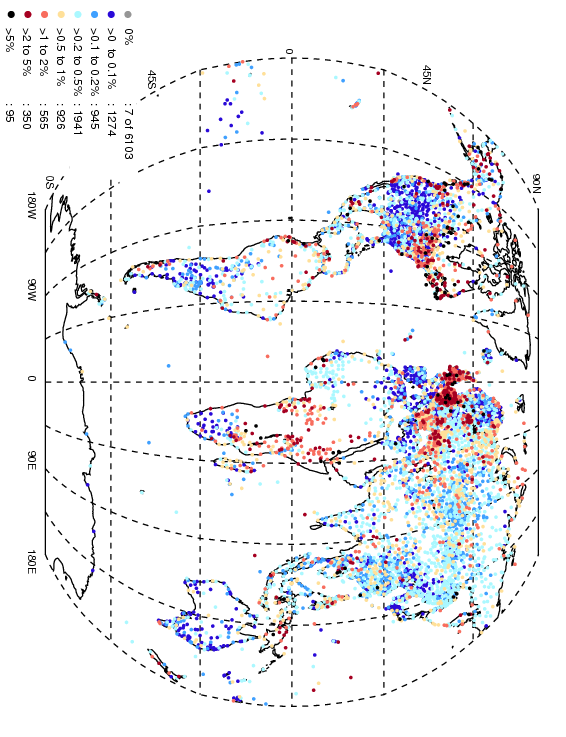}\\
\includegraphics[angle=90,width=8cm]{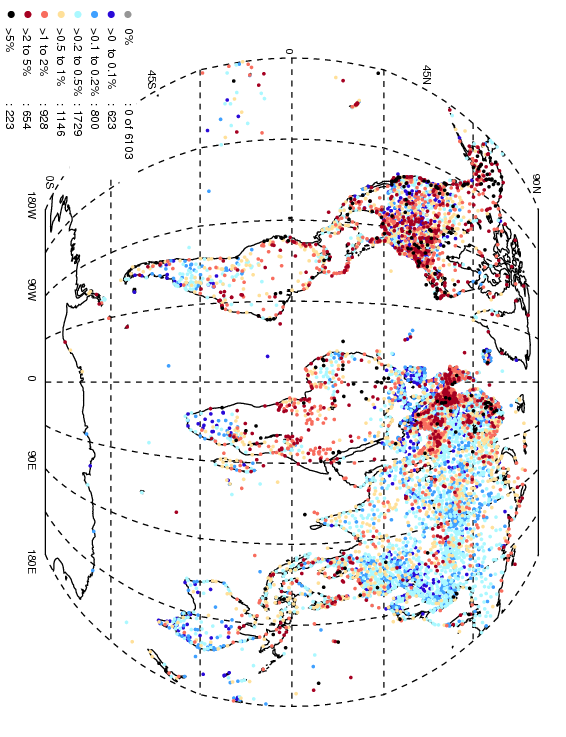}
\includegraphics[angle=90,width=8cm]{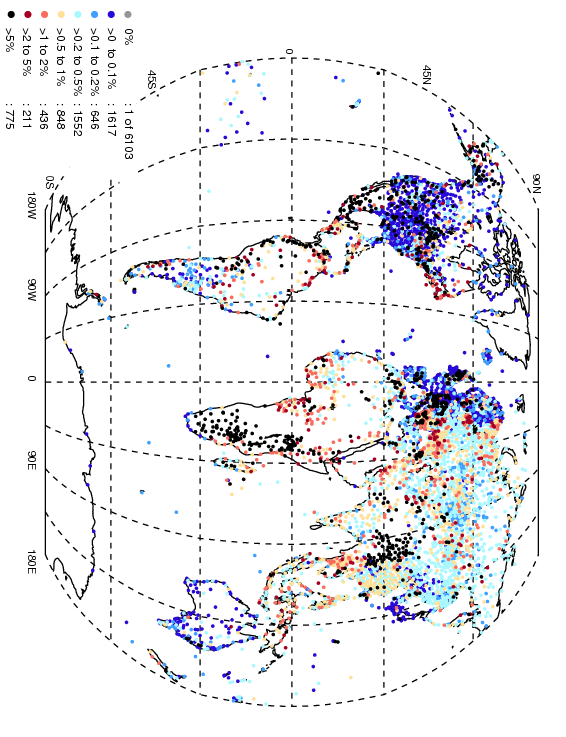}
\caption{\label{fig:19} Rejection rates by variable for each
  station. Top panel: temperature, Middle panel: dewpoint
  temperature, and Lower Panel: sea-level pressure. Different
  rejection rates are shown by different colours, and
  the key in each panel provides the total number of stations in
  each band.}
\end{figure}

Although a Gaussian fit may not be optimal or appropriate, it will
account for the spread of the majority of observations for each
station, and the contiguous portion of the distribution will be
retained. For Station 476960-43323 (Yokosuka, Japan, an assigned
composite) this part of the test flags a number of observations. In
fact, during the winter all temperature measurements below 0\,$^\circ$C appear
to be measured in Fahrenheit (see Fig.~\ref{fig:7})\footnote{Such an error has been noted and reported back to NCDC.}.  In months that have a
mixture of above and below 0\,$^\circ$C data (possibly Celsius and Fahrenheit
data), the monthly median may not show a large anomaly, so this
extension is needed to capture the bad data. Figure~\ref{fig:7} shows that the
two clusters of red points in January and October~1973 are captured by
this portion of the test.  By comparing the observations for a given
calendar month over all years, the difference between the two
populations is clear (see bottom panel in Fig.~\ref{fig:7a}).  If there are
two, approximately equally sized distributions in the station record,
then this test will not be able to choose between them.

\begin{figure}[t]
\includegraphics[width=8.5cm]{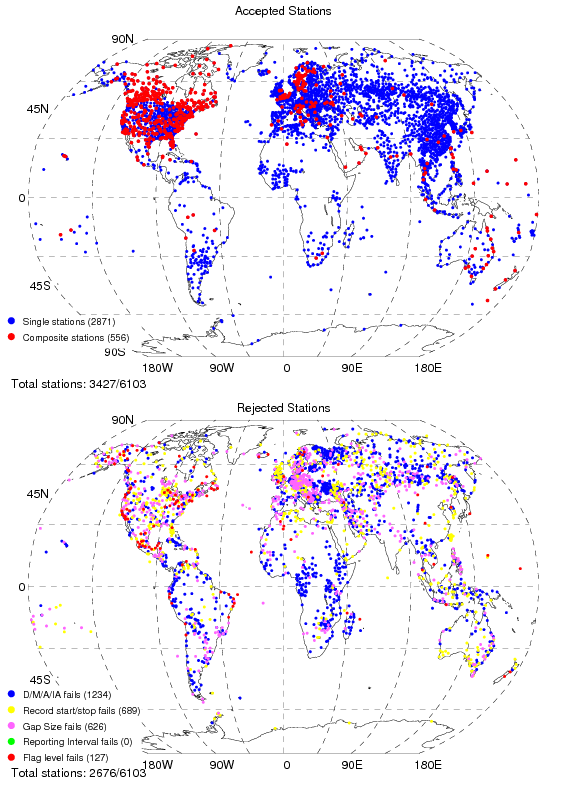}
\caption{\label{fig:20} The results of the final filtering to select
  climate quality stations.  Top:~the selected stations that pass the
  filtering, with red for composite stations
  (556/3427).  Bottom: the rejected stations.  Of these, 1234/2676
  fail to meet the daily, monthly, annual or interannual requirements (D/M/A/IA);
  689/2676 begin after 1980 or end before 2000;  626/2676 have a gap
  exceeding two years after the daily, monthly and annual completeness
  criteria have been applied; and 127 fail because one of the three
  main variables has a high proportion of flags.}
\end{figure}

To prevent the low pressure extremes associated with tropical cyclones
being excessively flagged, any low SLP observation identified by this
second part of the test is only tentatively flagged.  Simultaneous
wind speed observations, if present, are used to identify any storms
present, in which case low SLP anomalies are likely to be true. If the
simultaneous wind speed observations exceed the median wind speed for
that calendar month by 4.5 MADs, then \mbox{storminess} is \mbox{assumed} and the SLP
flags are unset.  If there are no wind data present, the neighbouring
stations can be used to unset these tentative flags in test 14.  The
tentative flags are only used for SLP observations in this test.

\subsubsection{Test 7: known records check}\label{sec:QC:tests:7}

Absolute limits are assigned based on recognised and documented world
and regional records (Table~\ref{table:3}). All hourly observations outside these
limits are flagged.  If temperature observations exceed a record, the
dewpoints are synergistically flagged.  Recent analyses of the record
Libyan temperature have resulted in a change to the global and African
temperature record \citep{ElFadli12}.  Any observations that would be
flagged using the new value but not by the old are likely to have been
flagged by another test.  This only affects African observations, and
those not assigned to the WMO regions outlined in
Table~\ref{table:3}.  The value used by this test will be updated in a
future release of HadISD.

\begin{figure*}[t]
\centering
\includegraphics[width=0.49\textwidth]{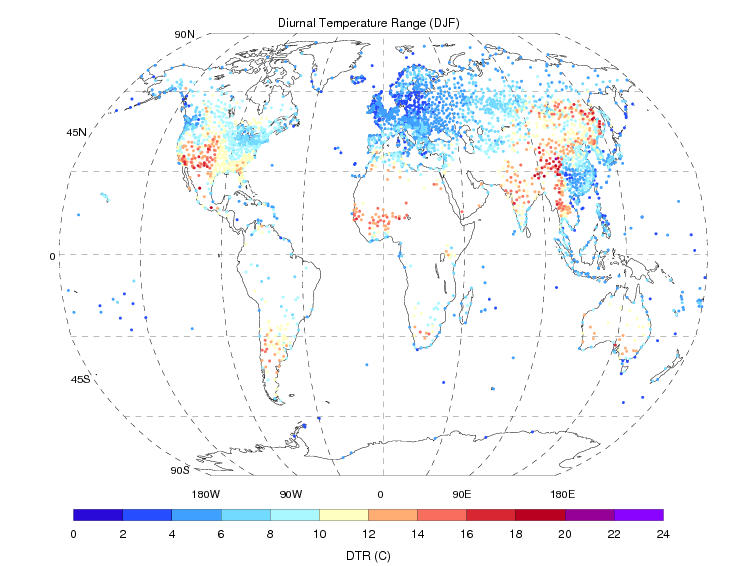}
\includegraphics[width=0.49\textwidth]{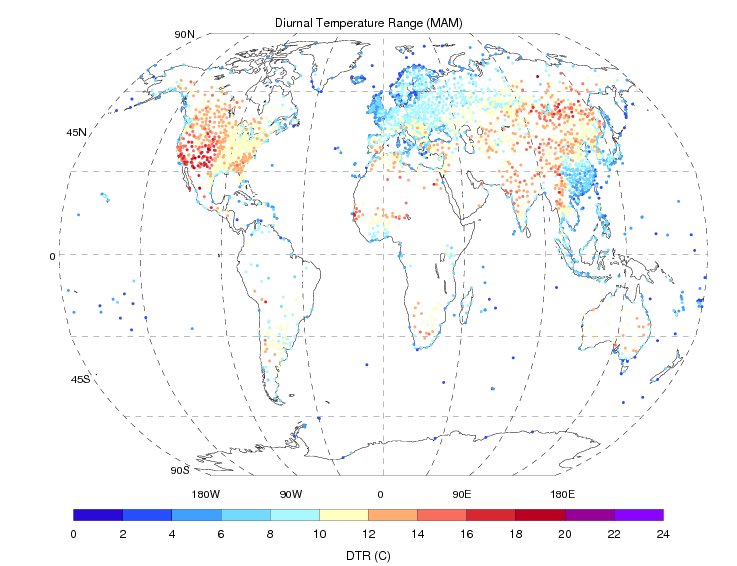}
\includegraphics[width=0.49\textwidth]{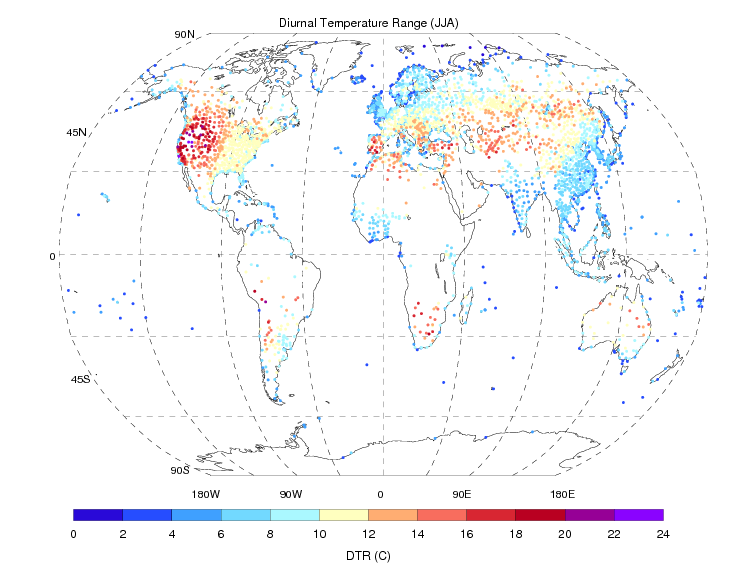}
\includegraphics[width=0.49\textwidth]{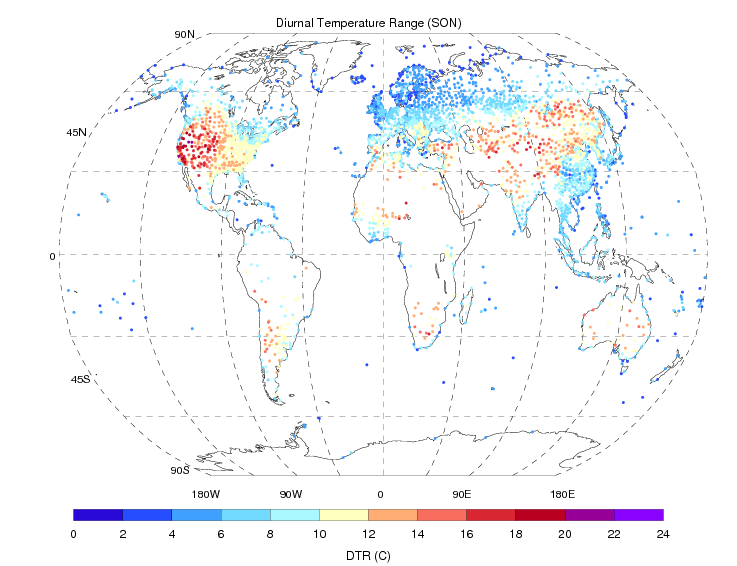}
\caption{\label{fig:21} The median diurnal temperature ranges recorded by
  each station (using the selected 3427 stations) for each of the four
  three-month seasons.  Top-left for December-January-February,
  top-right for March-April-May, bottom-left for June-July-August and
  bottom-right for September-October-November.}
\end{figure*}

\begin{figure*}[t]
\centering
\includegraphics[angle=90,width=7cm]{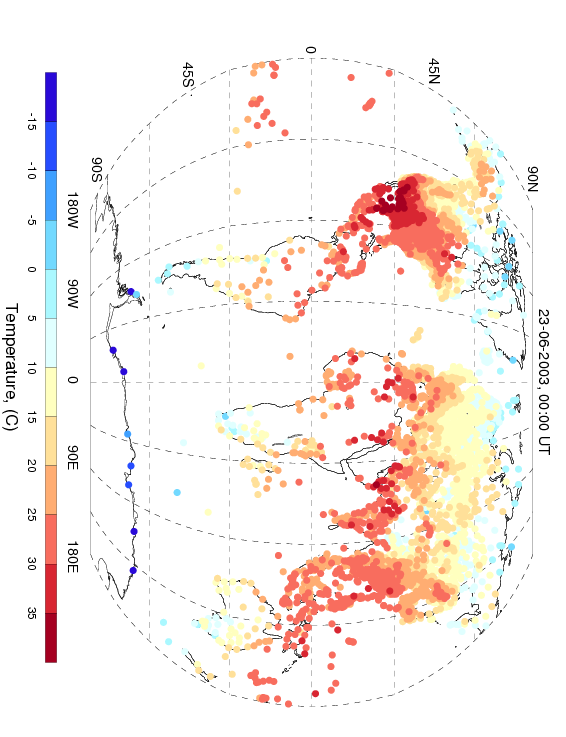}
\includegraphics[angle=90,width=7cm]{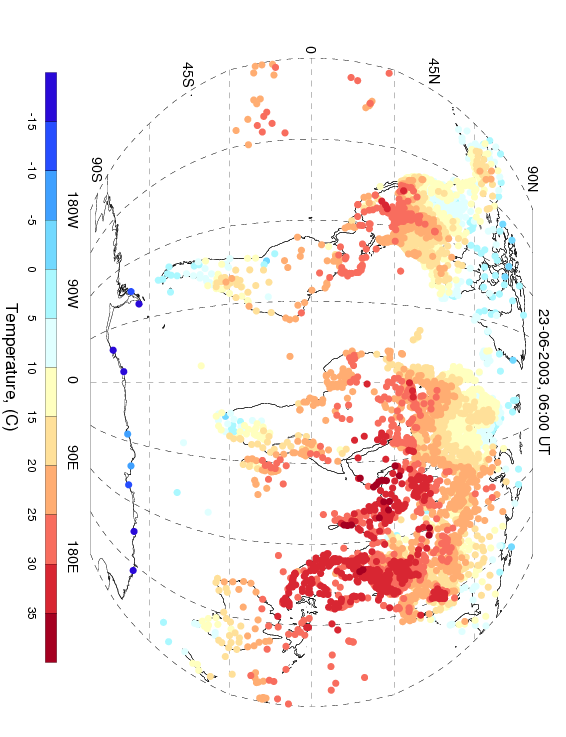}
\includegraphics[angle=90,width=7cm]{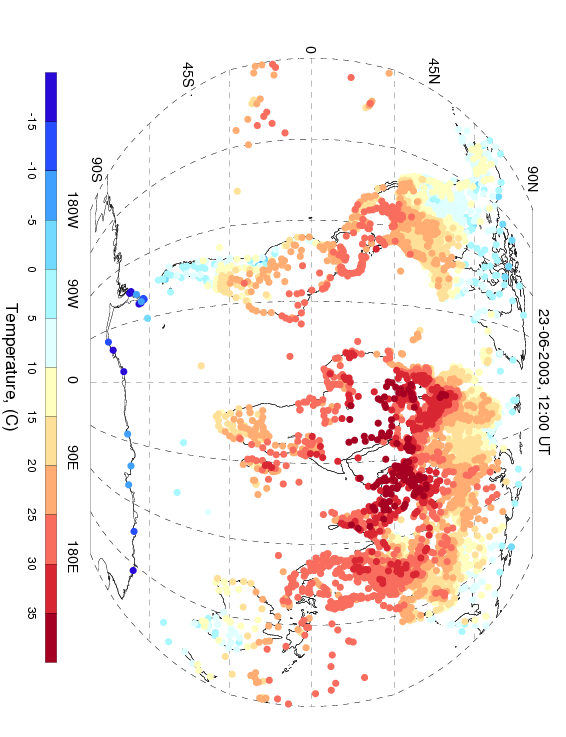}
\includegraphics[angle=90,width=7cm]{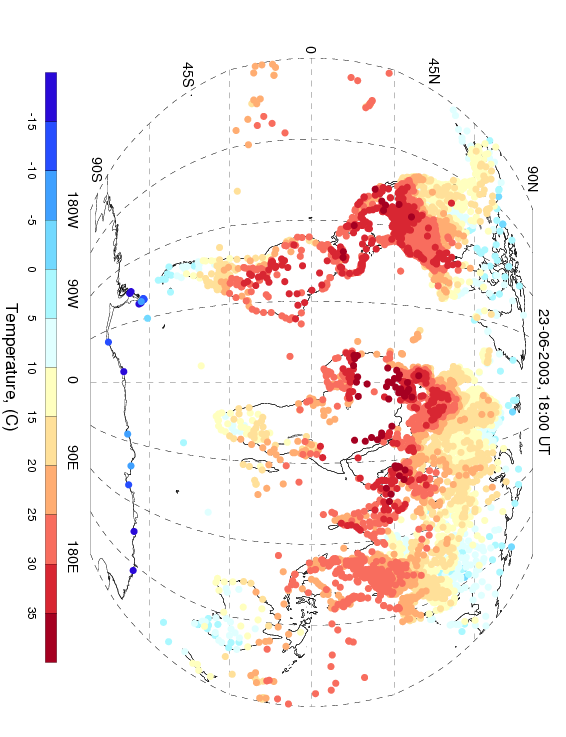}
\caption{\label{fig:22} The temperature for each station on
the 23~June~2003 at 00:00, 06:00, 12:00 and 18:00\,UT using all 6103
  stations.}
\end{figure*}

%t3
\begin{table*}[t]
\caption{\label{table:3}Extreme limits for observed variables gained
  from http://wmo.asu.edu (the official WMO climate extremes
  repository) and the GHCND tests. Dewpoint minima are estimates
  based upon the record temperature minimum for each region. First
  element in each cell is the minimum and the second the maximum legal
  value. Regions follow WMO regional definitions and are given at:
  http://weather.noaa.gov/tg/site.shtml. Global values are used for
  any station where the assigned WMO identifier is missing or does not
  fall within the region categorization. Wind speed and sea-level
  pressure records are not currently documented regionally so global
  values are used throughout.  We note that the value for the African
  and global maximum temperature has changed \citep{ElFadli12}.  This will be updated in
a future version of HadISD.}
\begin{tabular}{lccccccccccc}
\tophline
\raisebox{-3ex}[0mm][1mm]{Region}&\multicolumn{2}{c}{Temperature}&
&\multicolumn{2}{c}{Dewpoint Temperature}&&\multicolumn{2}{c}{Wind speed}&&\multicolumn{2}{c}{Sea-level pressure}\\
&\multicolumn{2}{c}{($^\circ$C)}&&\multicolumn{2}{c}{($^\circ$C)}&&\multicolumn{2}{c}{(m\,s$^{-1}$)}&&\multicolumn{2}{c}{(hPa)}\\
\cline{2-3}\cline{5-6}\cline{8-9}\cline{11-12}
&max&min&&max&min&&max&min&&max&min\\
\hhline
Global&$-$89.2&57.8&&$-$100.0&57.8&&0.0&113.3&&870&1083.3\\
Africa&$-$23.0&57.8&&$-$50.0&57.8&&--&--&&--&--\\
Asia&$-$67.8&53.9&&$-$100.0&53.9&&--&--&&--&--\\
S. America&$-$32.8&48.9&&$-$60.0&48.9&&--&--&&--&--\\
N. America&$-$63.0&56.7&&$-$100.0&56.7&&--&--&&--&--\\
Pacific&$-$23.0&50.7&&$-$50.0&50.7&&--&--&&--&--\\
Europe&$-$58.1&48.0&&$-$100.0&48.0&&--&--&&--&--\\
Antarctica&$-$89.2&15.0&&$-$100.0&15.0&&--&--&&--&--\\
\bottomhline
\end{tabular}
\end{table*}

\subsubsection{Test 8: repeated streaks/unusual spell frequency}\label{sec:QC:tests:8}

This test searches for consecutive observation replication, same hour
observation replication over, a number of days (either using a
threshold of a certain number of observations, or for sparser records,
a number of days during which all the observations have the same
value) and also whole day replication for a streak of days. All three
tests are conditional upon the typical reporting precision as coarser
precision reporting (e.g. temperatures only to the nearest whole
degree) will increase the chances of a streak arising by chance (Table~\ref{table:4}). For wind speed, all values below 0.5\,ms$^{-1}$ (or 1\,ms$^{-1}$ for coarse
recording resolution) are also discounted in the streak search given
that this variable is not normally distributed and there could be long
streaks of calm conditions.

%t4
\begin{table*}[t]
\caption{\label{table:4}Streak check criteria and their assigned
sensitivity to typical within-station reporting resolution for each
variable.}
\begin{tabular}{lrrrr}
\tophline
&Reporting&Straight repeat &Hour
repeat&Day repeat\\
Variable& resolution&streak criteria& streak criteria& streak criteria\\
\hhline
&1\,$^\circ$C& 40 values of 14 days & 25 days & 10 days\\
Temperature&0.5\,$^\circ$C& 30 values or 10 days & 20 days & 7 days\\
&0.1\,$^\circ$C& 24 values or 7 days & 15 days & 5 days\\
\hhline
&1\,$^\circ$C& 80 values of 14 days & 25 days & 10 days\\
Dewpoint&0.5\,$^\circ$C& 60 values or 10 days & 20 days & 7 days\\
&0.1\,$^\circ$C& 48 values or 7 days & 15 days & 5 days\\
\hhline
&1\,hPa& 120 values of 28 days & 25 days & 10 days\\
SLP&0.5\,hPa& 100 values or 21 days & 20 days & 7 days\\
&0.1\,hPa& 72 values or 14 days & 15 days & 5 days\\
\hhline
&1\,ms$^{-1}$& 40 values of 14 days & 25 days & 10 days\\
Wind speed&0.5\,ms$^{-1}$& 30 values or 10 days & 20 days & 7 days\\
&0.1\,ms$^{-1}$& 24 values or 7 days & 15 days & 5 days\\
\bottomhline
\end{tabular}
\vspace{5mm}
\end{table*}

During development of the test a number of station time series were
found to exhibit an alarming frequency of streaks shorter than the
assigned critical lengths in some years. An extra criterion was added
to flag all streaks in a given year when consecutive value streaks of
$>10$ elements occur with extraordinary frequency ($>5$ times the median
annual frequency). Station 724797-23176 (Milford, UT, USA, an assigned
composite) exhibits a propensity for streaks during 1981 and 1982 in
the dewpoint temperature (Fig.~\ref{fig:8}), which is not seen in any other
years or nearby stations.

\subsubsection{Test 9: climatological outlier
  check}\label{sec:QC:tests:9}

Individual gross outliers from the general station distribution are a
common error in observational data caused by random recording,
reporting, formatting or instrumental errors (Fiebrich and Crawford,
2009). This test uses individual observation deviations derived from
the monthly mean climatology calculated for each hour of the
day. These climatologies are calculated using observations that have
been winsorised\footnote{Winsorising is the process by which all
  values beyond a threshold value from the mean are set to that
  threshold value (5 and 95 per cent in this instance).  The number of
  data values in the population therefore remains the same, unlike
  trimming, where the data further from the mean are removed from the
  population \citep{Afifi79}.} to remove the initial effects of
outliers. The raw,
un-winsorised observations are anomalised using these climatologies
and standardised by the IQR for that month and hour. Values are
subsequently low-pass filtered to remove any climate change signal
that would cause overzealous removal at the ends of the time
series. In an analogous way to the distributional gap check, a
Gaussian is fitted to the histogram of these anomalies for each month,
and a threshold value, rounded outwards, is set where this crosses the
$y=0.1$ line. The distribution beyond this threshold value is scanned
for a gap (equal to the bin width or more), and all values beyond any
gap are flagged.  Observations that fall between the critical
threshold value and the gap or the critical threshold value and the
end of the distribution are tentatively flagged, as they fall outside
of the expected distribution (assuming it is Gaussian; see Fig.~\ref{fig:9}). These may be later reinstated on comparison with good data from
neighbouring stations (see Sect.~\ref{sec:QC:tests:14}). A caveat to protect low-variance
stations is added whereby the IQR cannot be less than 1.5\,$^\circ$C. When
applied to sea-level pressure, this test frequently flags storm
signals, which are likely to be of high interest to many users, and so
this test is not applied to the pressure data.

As for the distributional gap check, the Gaussian may not be the best
fit or even appropriate for the distribution, but by fitting to the
observed distribution, the spread of the majority of the observations
for the station is accounted for, and searching for a gap means that
the contiguous portion of \mbox{distribution} is retained.

\subsubsection{Test 10: spike check}\label{sec:QC:tests:10}

Unlike the operational ISD product, which uses a fixed value for all
stations (Lott et al., 2001), this test uses the filtered station time
series to decide what constitutes a ``spike'', given the statistics of
the series. This should avoid over zealous flagging of data in high
variance locations but at a potential cost for stations where false
data spikes are truly pervasive. A first difference series is created
from the filtered data for each time step (hourly, 2-hourly, 3-hourly)
where data exist within the past three hours. These differences for
each month over all years are then ranked and the IQR calculated.
Critical values of 6 times the rounded-up IQR are calculated for one-,
two- and three-hourly differences on a monthly basis to account for
large seasonal cycles in some regions. There is a caveat that no
critical value is smaller than 1\,$^\circ$C or hPa (conceivable in some
regions but below the typically expected reported resolution). Also
hourly critical values are compared with two hourly critical values to
ensure that hourly values are not less than 66 per cent of two hourly
values. Spikes of up to three sequential observations in the
unfiltered data are defined by satisfying the following criteria.  The
first difference change into the spike has to exceed the threshold and
then have a change out of the spike of the opposite sign and at least
half the critical amplitude. The first differences just outside of the
spike have to be under the critical values, and those within a
multi-observation spike have to be under half the critical value (see
Fig.~\ref{fig:10} highlighting the various thresholds).  These checks ensure
that noisy high variance stations are not overly flagged by this test.
Observations at the beginning or end of a contiguous set are also
checked for spikes by comparing against the median of the subsequent
or previous 10 observations. Spike check is particularly efficient at
flagging an apparently duplicate period of record for station
718936-99999 (Campbell River, Canada, an assigned composite station),
together with the climatological check~(Fig.~\ref{fig:11}).

\subsubsection{Test 11: temperature and dewpoint temperature
  cross-check}\label{sec:QC:tests:11}

Following \citep{Willett08}, this test is specific to humidity-related errors
and searches for three \mbox{different}~\mbox{scenarios:}
\begin{enumerate}
\item \textit{Supersaturation} (dewpoint temperature\,$>$\,temperature), although
physically plausible especially in very cold and humid climates
\citep{Makkonen05}, is highly unlikely in most
regions. Furthermore, standard meteorological instruments are
unreliable at measuring this accurately.
\item \textit{Wet-bulb reservoir drying} (due to evaporation or freezing) is very
common in all climates, especially in automated stations. It is
evidenced by extended periods of temperature equal to dewpoint
temperature (dewpoint depression of 0\,$^\circ$C).
\item \textit{Cutoffs of dewpoint temperatures at temperature extremes}.
Systematic flagging of dewpoint temperatures when the simultaneous
temperature exceeds a threshold (specific to individual National
Meteorological Services' recording methods) has been a common practice
historically with radiosondes \citep{Elliott95,McCarthy09}. This has
also been found in surface stations both for hot and cold extremes
\citep{Willett08}.

\end{enumerate}

For supersaturation, only the dewpoint temperature is flagged if the
dewpoint temperature exceeds the temperature. The temperature data may
still be desirable for some users. However, if this occurs for 20 per
cent or more of the data within a month, then the whole month is
flagged. In fact, no values are flagged by this test and a later, independent check
run at NCDC showed that there were no episodes of supersaturation in
the raw ISD (Neal Lott, personal communication).  However it is retained
for completeness. For wet-bulb reservoir drying, all continuous
streaks of absolute dewpoint depression $<0.25$\,$^\circ$C are noted. The leeway
of $\pm0.25$\,$^\circ$C allows for small systematic differences between the
thermometers. If a streak is $>24$\,h with $\geq$ four observations
present, then all the observations of dewpoint temperature are flagged
unless there are simultaneous precipitation or fog observations for
more than one-third of the continuous streak. We use a cloud base
measurement of $<1000$ feet to indicate fog as well as the present
weather information. This attempts to avoid over zealous flagging in
fog- or rain-prone regions (which would dry-bias the observations if
many fog or rain events were removed). However, it is not perfect as
not all stations include these variables. For cutoffs, all
observations within a month are binned into 10\,$^\circ$C temperature bins from
$-$90\,$^\circ$C to 70\,$^\circ$C (a range that extends wider than recognised historically
recorded global extremes). For any month where at least 50 per cent of
temperatures within a bin do not have a simultaneous dewpoint
temperature, all temperature and dewpoint data within the bin are
flagged. Reporting frequencies of temperature and dewpoint are
identified for the month, and removals are not applied where
frequencies differ significantly between the variables.  The cutoffs
part of this test can flag good dewpoint data even if only a small
portion of the month has problems, or if there are gaps in the dewpoint
series that are not present in the temperature observations.

\subsubsection{Test 12: cloud coverage logical checks}\label{sec:QC:tests:12}

Synoptic cloud data are a priori a very difficult parameter to test
for quality and homogeneity. Traditionally, cloud base height and
coverage of each layer (low, mid, and high) in oktas were estimated
by eye. Now cloud is observed in many countries primarily using a
ceilometer which takes a single 180$^\circ$ scan across the sky with a very
narrow off-scan field-of-view. Depending on cloud type and cloud
orientation, this could easily under- or over-estimate actual sky
coverage. Worse, most ceilometers can only observe low or at best
\mbox{mid-level} clouds. Here, a conservative approach has been taken where
simple cross checking on cloud layer totals is used to infer basic
data quality. This should flag the most glaring issues but does not
guarantee a high quality database.

Six tests are applied to the data. If coverage at any level is given
as 9 or 10, which officially mean sky obscured and partial obstruction
respectively, that individual value is flagged\footnote{All ISD values
  greater than 10, which signify scattered, broken and full cloud for
  11, 12 and 13 respectively, have been converted to 2, 4 and 8 oktas
  respectively during {netCDF} conversion prior to QC.}. If total cloud cover
is less than the sum of low, middle and high level cloud cover, then
all are flagged. If low cloud is given as 8 oktas (full coverage) but
middle or high level clouds have a value, then, as it is not
immediately apparent which observations are at fault, the low, middle
and/or high cloud cover values are flagged. If middle layer cloud is
given as 8 oktas (full coverage) but high level clouds have a value,
then, similarly, both the middle and high cloud cover value are flagged. If
the cloud base height is given as 22\,000, this means that the cloud
base is unobservable (sky is clear). This value is then set to $-$10 for
computational reasons.  Finally, cloud coverage can only be from 0 to
8 oktas. Any value of total, low, middle layer or high cloud that is
outside these bounds is flagged.

\subsubsection{Test 13: unusual variance check}\label{sec:QC:tests:13}

The variance check flags whole months of temperature, dewpoint
temperature and sea-level pressure where the within month variance of
the normalised anomalies (as described for climatological check) is
sufficiently greater than the median variance over the full station
series for that month based on winsorised data \citep{Afifi79}. The
variance is taken as the MAD of the normalised anomalies in
each individual month with $\geq120$ observations. Where there is
sufficient representation of that calendar month within the time
series (10 months each with $\geq120$ observations), a median variance
and IQR of the variances are calculated. Months that differ by more
than 8 IQR (temperatures and dewpoints) or 6 IQR (sea-level pressures)
from the station month median are flagged. This threshold is increased
to 10 or 8 IQR respectively if there is a reduction in reporting
frequency or resolution for the month relative to the majority of the
time series.

Sea-level pressure is accorded special treatment to reduce the removal
of storm signals (extreme low pressure). The first
difference series is taken. Any month where the largest consecutive
negative or positive streak in the difference series exceeds 10 data
points is not considered for removal as this identifies a spike in the
data that is progressive rather than transient. Where possible, the
wind speed data are also included, and the median found for a given
month over all years of data.  The presence of a storm is determined
from the wind speed data in combination with the sea-level pressure
profile.  When the wind speed climbs above 4.5 MADs from the median
wind speed value for that month and if this peak is coincident with a
minimum of the sea-level pressure ($\pm24$\,h), which is also more than
4.5 MADs from the median pressure for that month, then storminess is
assumed.  If these criteria are satisfied, then no flag is set.  This
test for storminess includes an additional test for unusually low SLP
values, as initially this QC test only identifies periods of high
variance.  Figure~\ref{fig:12}, for station 912180-99999 (Andersen Air Force Base,
Guam), illustrates how this check is flagging obviously dubious
dewpoints that previous tests had failed to identify.

\subsubsection{Test 14: nearest neighbour data checks}\label{sec:QC:tests:14}

Recording, reporting or instrument error is unlikely to be replicated
across networks. Such an error may not be detectable from the
intra-station distribution, which is inherently quite noisy. However,
it may stand out against simultaneous neighbour observations if the
correlation decay distance \citep{Briffa93} is large compared to
the actual distance between stations and therefore the noise in the
difference series is comparatively low. This is usually true for
temperature, dewpoint and pressure. However the check is less powerful
for localised features such as convective \mbox{precipitation} or storms.

%t5
\begin{table*}[p]
\caption{\label{table:5}Summary of tests applied to the data.}
\scalebox{.84}[.84]{\begin{tabular}{p{4.4cm}ccccccp{5cm}p{5cm}}
\tophline
Test&\multicolumn{6}{c}{Applies to}&Test failure& Notes        \\
\cline{2-7}
(Number)&T & Td  &SLP &ws  &wd &clouds&  criteria  \\
\hhline
\multicolumn{9}{c}{Intra-station}\\
\hhline
Duplicate months check (2)&X & X   & X  & X  &X  &   X  & Complete
match to least temporally complete month's record for T\\
\hline
Odd cluster check (3)&X & X   & X  & X  &X  &   & $\leq 6$ values in 24\,h separated from any other data by $>48$\,h&Wind direction
removed using wind speed characteristics\\
\hhline
Frequent values check (4)&X & X   & X  &    &    &    & Initially $>50$\,\%
of all data in current 0.5\,$^\circ$C or hPa bin out of this and $\pm3$ bins for
all data to highlight, with $\geq30$ in the bin.  Then on yearly basis
using highlighted bins with $>50$\,\% of data and $\geq20$ observations in
this and $\pm3$ bins OR $>90$\,\% data and $\geq10$ observations in this and
$\pm3$ bins.  For seasons, the bin size thresholds are reduced to 20, 15
and 10 respectively. &Histogram approach for computational expediency.
T and Td synergistically removed, if T is bad, then Td is removed and
vice versa.\\
\hhline
Diurnal cycle check (5)&X & X   & X  & X  &X  &   X  & 30 days without
3 consecutive good fit/missing or 6 days mix of these to T diurnal
cycle.\\
\hhline
Distributional gap check (6)&X & X   & X  &    &       &    & Monthly
median anomaly $>5$IQR from median.  Monthly median anomaly at least
twice the distance from the median as the other tail and $>1.5$
IQR. Data outside of the Gaussian distribution for each calendar month
over all years, separated from the main population.&All months in tail
with apparent gap in the distribution are removed beyond the assigned
gap for the variable in question. Using the distribution for all
calendar months, tentative flags set if further from mean than
threshold value.  To keep storms, low SLP observations are only
tentatively\\
\hhline
Known record check (7)&X & X   & X  & X  &       &    & See  Table \ref{table:3}   &Td
flagged if T flagged.   \\
\hhline
Repeated streaks/unusual spell \par frequency check (8)&X & X   & X  & X  &
&     & See Table \ref{table:4}  \\
\hhline
Climatological outliers check (9)&X & X   &     &    &       &    & Distribution
of normalised (by IQR) anomalies investigated for outliers using same
method as for distributional gap test.&To keep low variance stations,\par
minimum IQR is 1.5\,$^\circ$C\\
\hhline
Spike check (10)&X & X   & X  &    &       &    & Spikes of up to 3
consecutive points allowed.  Critical value of 6IQR (minimum 1\,$^\circ$C or
hPa) of first difference at start of spike, at least half as large and
in opposite direction at end.&First differences outside and inside  a
spike have to be under the critical and half the critical values
respectively\\
\hhline
T and Td cross-check: Supersaturation (11)&  & X   &  &    &       &    &
Td\,$>$\,T&Both variables removed, all data removed for a month if $>20$\,\% of
data fails\\
\hhline
T and Td cross-check: Wet bulb drying (11)&  & X   &  &    &       &    &
T\,$=$\,Td $>24$\,h and $>4$ observations unless rain / fog (low cloud
base) reported for $> 1/3$ of string&0.25\,$^\circ$C leeway allowed.\\
\hhline
T and Td cross-check: Wet bulb cutoffs (11)&  & X   & &    &       &    &
$>20$\,\% of T has no Td within a 10\,$^\circ$C T bin&Takes into account that Td at
many stations reported less frequently than T.\\
\hhline
Cloud coverage logical checks (12)&  &       &  &    &       &   X  &
Simple logical criteria (see Sect.~\ref{sec:QC:tests:12})\\
\hhline
Unusual variance check (13)&X & X   & X  &    &  &    & Winsorised
normalised (by IQR) anomalies exceeding 6~IQR after filtering&8 IQR if
there is a change in reporting frequency or resolution.  For SLP first
difference series used to find spikes (storms).  Wind speed also used
to identify storms\\
\bottomhline
\end{tabular}}
\end{table*}

%t5
\addtocounter{table}{-1}
\begin{table*}[t]
\caption{\label{table:5}Continued.}
\scalebox{.84}[.84]{\begin{tabular}{p{4.4cm}ccccccp{5cm}p{5cm}}
\tophline
Test&\multicolumn{6}{c}{Applies to}&Test failure& Notes        \\
\cline{2-7}
(Number)&T & Td  &SLP &ws  &wd &clouds&  criteria  \\
\hhline
\multicolumn{9}{c}{Intra-station}\\
\hhline
Inter-station duplicate check (1)&X &        &  &    &       &    & $>1000$
valid points and $>25$\,\% exact match over $t-11$ to $t+11$ window, followed
by manual assessment of identified series&Stations identified as
duplicates \par removed in entirety.\\
\hhline
Nearest neighbour data check (14)&X & X   & X  &    &  &    & $>2/3$ of
station comparisons suggest the value is anomalous within the
difference series at the 5 IQR level.&At least three and up to ten
neighbours within 300\,km and 500\,m, with preference given to filling
directional quadrants over distance in neighbour selection. Pressure
has additional caveat to ensure against removal of severe storms.\\
\hhline
Station clean-up (15)&X & X   & X  & X  &X  &   & $< 20$ values per
month or $>40$\,\% of values in a given month flagged for the
variable&Results in removal of whole month for that variable\\
\bottomhline
\end{tabular}}
\end{table*}

For each station, up to ten nearest neighbours (within 500\,m elevation
and 300\,km distance) are identified. Where possible, all four quadrants
(northeast, southeast, southwest and northwest) surrounding the
station must be represented by at least two neighbours to prevent
geographical biases arising in areas of substantial gradients such as
frontal regions. Where there are less than three valid neighbours, the
nearest neighbour check is not applied. In such cases the station ID
is noted, and these stations can be found on the HadISD website.  The
station may be of questionable value in any subsequent homogenisation
procedure that uses neighbour comparisons. A difference series is
created for each candidate station minus neighbour pair. Any
observation associated with a difference exceeding 5IQR of the whole
difference series is flagged as potentially dubious. For each time
step, if the ratio of dubious candidate-neighbour differences flagged
to candidate-neighbour differences present exceeds 0.67 (2 in 3
comparisons yield a dubious value), and there are three or more
neighbours present, then the candidate observation differs
substantially from most of its neighbours and is flagged.
Observations where there are fewer than three neighbours that have
valid data are noted in the flag array.

For sea-level pressure in the tropics, this check would remove some
negative spikes which are real storms as the low pressure core can be
narrow. So, any candidate-neighbour pair with a distance greater than
100\,km between is assessed. If 2/3 or more of the difference series
flags (over the entire record) are negative (indicating that this site
is liable to be affected by tropical storms), then only the positive
differences are counted towards the potential neighbour outlier
removals when all neighbours are combined. This succeeds in retaining
many storm signals in the record. However, very large negative spikes
in sea-level pressure (tropical storms) at coastal stations may still
be prone to removal especially just after landfall in relatively
station dense regions (see Sect.~\ref{sec:valid:katrina}). Here, station distances may
not be large enough to switch off the negative difference flags but
distant enough to experience large differences as the storm
passes. Isolated island stations are not as susceptible to this
effect, as only the station in question will be within the
low-pressure core and the switch off of negative difference flags will
be activated. Station 912180-99999 (Anderson, Guam) in the western
Tropical Pacific has many storm signals in the sea-level pressure
(Fig.~\ref{fig:13}). It is important that these extremes are not removed.

Flags from the spike, gap (tentative low SLP flags only; see Sect.~\ref{sec:QC:tests:6}),
climatological (tentative flags only; see Sect.~\ref{sec:QC:tests:9}), odd cluster and
dewpoint depression tests (test numbers 3, 6, 9, 10 \& 11) can be unset
by the nearest neighbour data check.  For the first four tests this
occurs if there are three or more neighbouring stations that have
simultaneous observations that have not been flagged.  If the
difference between the observation for the station in question and the
median of the simultaneous neighbouring observations is less than the
threshold value of 4.5 MADs\footnote{As calculated from the
  neighbours observations, \mbox{approximately}~3$\sigma$.}, then the flag
is removed.  These
criteria are to ensure that only observations that are likely to be
good can have their flags removed.

In cases where there are few neighbouring stations with unflagged
observations, their distribution can be very narrow.  This narrow
distribution, when combined with poor instrumental reporting accuracy,
can lead to an artificially small MAD, and so to the erroneous
retention of flags.  Therefore, the MAD is restricted to a minimum of
0.5 times the worst reporting accuracy of all the stations involved
with this test.  So, for example, for a station where one neighbour
has 1\,$^\circ$C reporting, the threshold value is 2.25\,$^\circ$C\,=\,0.5\,$\times$\,1\,$^\circ$C\,$\times$\,4.5.

Wet-bulb reservoir drying flags can also be unset if more than two-thirds
of the neighbours also have that flag set. Reservoir drying
should be an isolated event, and so simultaneous flagging across
stations suggests an actual high humidity event. The tentative
climatological flags are also unset if there are insufficient
neighbours.  As these flags are only tentative, without sufficient
neighbours there can be no definitive indication that the observations
are bad, and so they need to be retained.

\subsubsection{Test 15: station clean-up}\label{sec:QC:tests:15}

A final test is applied to remove data for any month where there are
$<20$ observations remaining or $>40$ per cent of observations removed by
the QC. This check is not applied to cloud data as errors in cloud
data are most likely due to isolated manual errors.

\subsection{Test order}\label{sec:QC:order}

The order of the tests has been chosen both for computational
convenience (intra-station checks taking place before inter-station
checks) and also so that the most glaring errors are removed early on
such that distributional checks (which are based on observations that
have been filtered according the flags set thus far) are not biased.
Inter-station duplicate check (test 1) is run only once, followed by
the latitude and longitude check.  Tests 2 to 13 are run through in
sequence followed by test 14, the neighbour check.  At this point the
flags are applied creating a masked, preliminary, quality-controlled
dataset, and the flagged values copied to a separate store in case any
user wishes to retrieve them at a later date.  In the main data stream
these flagged observations are marked with a flagged data indicator,
different from the \mbox{missing} data indicator.

Then the spike (test 10) and odd-cluster (test 3) tests are re-run on
this masked data.  New spikes may be found using the masked data to
set the threshold values, and odd clusters may have been left after
the removal of bad data.  Test 14 is re-run to assess any further
changes and reinstate any  tentative flags  from the rerun of tests 3
and 10 where appropriate.  Then the clean-up of bad months, test 15,
is run and the flags applied as above creating a final
quality-controlled dataset.  A simple flow diagram is shown in Fig.~\ref{fig:3} indicating
the order in which the tests are applied. Table~\ref{table:5}
summarises which tests are applied to which data, what critical values
were applied, and any other relevant notes. Although the final quality-controlled
suite includes wind speed, direction and cloud data, the
tests concentrate upon SLP, temperature and dewpoint temperature and
it is these data that therefore are likely to have the highest
quality; so users of the remaining variables should take great
care. The typical reporting resolution and frequency are also
extracted and stored in the output {netCDF} file header fields.

%t6
\begin{table*}[t]
\caption{\label{table:6}Summary of removal of data from individual
  stations by the different tests for the 6103 stations considered in
  detailed analysis. The final column denotes any geographical
  prevalence. A version of this table in percent is presented in
  Table~\ref{table:9}.}
\scalebox{.84}[.84]{\begin{tabular}{p{3.6cm}lllllllllp{5cm}}
\tophline
Test&Variable      &\multicolumn{8}{c}{Stations with detection rate band (\% of total original observations)}& Notes on geographical prevalence \\
\cline{3-10}
(Number)&          &0          &0--0.1          &0.1--0.2        &0.2--0.5  &0.5--1.0      &1.0--2.0      &2.0--5.0        &5.0& of extreme removals \\
\hhline
Duplicate months check (2)&All             &6103         &0      &0        &0        &0          &0        &0        &0 &  \\
\hhline
Odd cluster check (3)&T        &2041         &2789    &484     &413      &213      &126        &34       &3 &  Ethiopia,
 Cameroon, Uganda,  Ukraine, Baltic states, Pacific coast of Colombia, Indonesian Guinea\\
&Td          &1855         &2946     &485    &439      &214       &128     &35         &1 &  As for temperature\\
&SLP         &1586         &3149     &567    &487      &203      &94         &17       &0 &  Cameroon, Ukraine,
Bulgaria, Baltic states, Indonesian Guinea\\
&ws          &1959         &2851     &480    &435      &218       &125     &32         &3 &  As for temperature\\
\hhline
Frequent values check (4)&T          &5980         &93           &14       &7        &4
&3        &2        &0 &  Largely random. Generally more prevalent in tropics, particularly Kenya\\
&Td          &5941         &91           &19       &17       &14       &8          &8        &5 &  Largely random, particularly bad in Sahel region and Philippines.\\
&SLP         &5998         &27           &8        &8        &9          &4        &31       &18  &  Almost exclusively Mexican stations.  Also a few UK stations.\\
\hhline
Diurnal cycle check (5)&All          &5765         &1      &16       &179      &70       &35         &24       &12  &  Mainly NE N. America, central Canada and central Russia regions\\
\hhline
Distributional gap check (6)&T             &2570         &3253     &42     &81       &77       &33         &38       &9 &  Mainly mid- to high-latitudes, more in N. America and central Asia\\
&Td          &1155         &4204     &298    &245      &114       &45        &36       &6 &  Mainly mid- to high-latitudes, more in N. America and central Asia\\
&SLP         &2736         &3096     &73     &90       &55       &28         &18       &7 &  Scattered\\
\hhline
Known records check (7)&T            &5313         &785          &1        &4        &0          &0        &0        &0 &  S. America, central Europe\\
&Td          &6090         &12           &0        &1        &0          &0        &0        &0 &  \\
&SLP         &4872         &1228     &2      &1        &0          &0        &0        &0 &  Worldwide apart from N. America, Australia, E China, Scandinavia\\
&ws          &6103         &0      &0        &0        &0          &0        &0        &0 &  \\
\bottomhline
\end{tabular}}
\end{table*}

%t6
\addtocounter{table}{-1}
%t6
\begin{table*}[t]
\caption{\label{table:6}Continued.}
\scalebox{.84}[.84]{\begin{tabular}{p{3.2cm}p{1.3cm}llllllllp{5cm}}
\tophline
Test&Variable      &\multicolumn{8}{c}{Stations with detection rate band (\% of total original observations)}& Notes on geographical prevalence \\
\cline{3-10}
(Number)&          &0          &0--0.1          &0.1--0.2        &0.2--0.5  &0.5--1.0      &1.0--2.0      &2.0--5.0        &$<5.0$& of extreme removals \\
\hhline
Repeated streaks/unusual spell frequency check (8)&T         &4785         &259          &183      &284      &218
&212     &146        &16  &  Particularly Germany, Japan, UK, Finland and NE. America and Pacific Canada/Alaska\\
&Td          &4343         &220          &190      &349      &344       &353     &272        &32  &  Similar to T, but more prevalent, additional cluster in Caribbean.\\
&SLP         &5996         &26           &13       &15       &8          &7        &22       &16  &  Almost exclusively Mexican stations\\
&ws          &5414         &243          &147      &135      &74       &60         &23       &7 &  Central \& northern South America, Eastern Africa, SE Europe, S Asia, Mongolia\\
\hhline
Climatological outliers \par check (9)              &T            &1295         &4382     &217    &159      &36       &13         &1        &0 &  Fairly uniform, but higher in tropics\\
&Td          &1064         &4538     &238    &192      &49       &19         &3        &0 &  As for temperature\\
\hhline
Spike check (10)&T             &95         &3650     &1270     &992      &92       &2          &2        &0 &  Fairly uniform, higher in Asia, especially eastern China\\
&Td          &38         &3567     &1486     &940      &66       &4          &2        &0 &  As for temperature\\
&SLP         &760        &3437     &1068     &802      &33       &3          &0        &0 &  Fairly uniform, few flags in southern Africa and western China\\
\hhline
T and Td cross-check: Supersaturation (11)&T, Td       &6103         &0      &0        &0        &0          &0        &0        &0 &  \\
\hhline
T and Td cross-check: Wet bulb drying (11)&Td          &3982         &1721     &194    &140      &37
&22         &6        &1 &  Almost exclusively NH extra-tropical, concentrations, Russian high Arctic, Scandinavia, Romania.\\
\hhline
T and Td cross-check: Wet bulb cutoffs (11)    &Td           &5055         &114          &211      &319
&175      &128        &69       &32  &  Mainly high latitude/elevation stations, particularly Scandinavia, Alaska, Mongolia, Algeria, USA.\\
\hhline
Cloud coverage logical check (12)&Cloud variables  &1    &682          &471      &979      &1124     &1357     &1173
&315 &  Worst in Central/Eastern Europe, Russian and Chinese coastal sites, USA, Mexico, eastern central Africa.\\
\hhline
Unusual variance check (13)&T        &5658         &13           &80       &298      &50       &4          &0
&0 &  Most prevalent in parts of Europe, US Gulf and west coasts\\
&Td          &5605         &13           &78       &334      &60       &10         &3        &0 &
Largely Europe, SE Asia and Caribbean/Gulf of Mexico\\
&SLP         &5263         &33           &118      &494      &150       &26        &10       &9 &
Almost exclusively tropics, particularly prevalent in sub-Saharan Africa, Ukraine, also eastern China\\
\hhline
Nearest neighbour data check (14)&T        &1549         &4369     &94     &35       &27       &19         &12       &1 &  Fairly uniform, worst in Ukraine, UK, Alaska\\
&Td          &1456         &4368     &159    &58       &36       &16         &10       &0 &  As for temperature\\
&SLP         &1823         &3995     &203    &58       &13       &4          &6        &1 &  Fairly uniform, worst in Ukraine, UK, eastern Arctic Russia\\
\hhline
Station clean-up (15)&T        &3865         &1546     &239    &219      &138       &60        &27       &9 &  High latitude N. America, Vietnam, eastern Europe, Siberia\\
&Td          &3756         &1526     &240    &277      &159       &85        &46       &14  &  Very similar to that for temperatures\\
&SLP      &3244         &2242     &212       &224      &107       &40        &29       &5 &  Many in Central America, Vietnam, Baltic states. \\
&ws       &3900         &1584     &214       &226      &108       &40        &24       &7 &  Western tropical coasts of Central America, central \& eastern Africa, Myanmar, Indonesia\\
\bottomhline
\end{tabular}}
\end{table*}

\subsection{Fine-tuning}\label{sec:QC:fineTuning}

In order to fine-tune the tests and their critical and threshold
values, the entire suite was first tested on the 167 stations in the
British Isles.  To ensure that the tests were still capturing known
and well-documented extremes, three such events were studied in
detail: the European heat wave in August~2003 and the storms of
October~1987 and January~1990.  During the course of these analyses, it
was noted that the tests (in their then current version) were not
performing as expected and were removing true extreme values as
documented in official Met Office records and literature for those
events.  This led to further fine-tuning and additions resulting in
the tests as presented above.  All analyses and diagrams are from the
quality control procedure after the updates from this fine-tuning.

As an example Fig.~\ref{fig:14} shows the passage of the low pressure core of
the 1987 storm.  The low pressure minimum is clearly not excluded by
the tests as they now stand, whereas previously a large number of valid
observations around the low pressure minimum were flagged.  The two
removed observations come from a single station and were flagged by
the spike test (they are clear anomalies above the remaining SLP
observations; see Fig.~\ref{fig:15}).

Any pervasive issues with the data or individual stations will be
reported to the ISD team at NCDC to allow for the improvement of
the data for all users.  We encourage users of HadISD who discover suspect
data in the product to contact the authors to allow the station to be
investigated and any improvements to the raw data or the QC suite to
be applied.

NCDC provide a list of known issues with the ISD database (\url{http://www1.ncdc.noaa.gov/pub/data/ish/isd-problems.pdf}).  Of the 27
problems known at the time of writing (31~July~2012), most are for
stations, variables or time periods which are not included in the
above study. Of the four that relate to data issues that could be
captured by the present analysis, all the bad data were successfully identified and removed
(numbers 6, 7, 8 and 25, stations 718790, 722053, 722051 and 722010).
Number 22 has been solved during the compositing process (our station
725765-24061 contains both 725765-99999 and 726720-99999).  However,
number 24 (station 725020-14734) cannot be detected by the
QC suite as this error relates to the reporting accuracy of the instrument.

\section{Validation and analysis of quality control results}\label{sec:validation}

To determine how well the dataset captures extremes, a number of known
extreme climate events from around the globe were studied to determine
the success of the QC procedure in retaining extreme values while
removing bad data.  This also allows the limitations of the QC
procedure to be assessed.  It also ensures that the fine-tuning
outlined in Sect.~4.2 did not lead to at least gross over-tuning
being based upon the climatic characteristics of a single relatively
small region of~the~globe.

\subsection{Hurricane Katrina, September 2005}\label{sec:valid:katrina}

Katrina formed over the Bahamas on 23~August~2005 and crossed
southern Florida as a moderate Category 1 hurricane, causing some
deaths and flooding.  It rapidly strengthened in the Gulf of Mexico,
reaching Category 5 within a few hours. The storm weakened before
making its second \mbox{landfall} as a Category 3 storm in southeast
Louisiana. It was one of the strongest storms to hit the USA, with
sustained winds of 127\,mph at landfall, equivalent to a Category 3
storm on the Saffir-Simpson scale \citep{Graumann06}.  After
causing over \$100  billion of damage and 1800 deaths in Mississippi
and Louisiana, the core moved northwards before being \mbox{absorbed} into a
front around the Great Lakes.

Figure~\ref{fig:16} shows the passage of the low pressure core of Katrina over
the southern part of the USA on 29 and 30~August~2005.  This
passage can clearly be tracked across the country.  There are a number
of observations which have been removed by the QC, highlighted in the
figure.  These observations have been removed by the neighbour check.
This identifies the issue raised in Sect.~\ref{sec:QC:tests:14} (test 14), where even
stations close by can experience very different simultaneous sea-level
pressures with the passing of very strong storms. However the passage
of this pressure system can still be \mbox{characterised} from this dataset.

\subsection{Alaskan cold spell, February~1989}\label{sec:valid:alaska}

The last two weeks of January~1989 were extremely cold throughout
Alaska except the Alaska Panhandle and Aleutian Islands.  A number of new
minimum temperature records were set (e.g.~$-$60.0\,$^\circ$C at Tanana and
$-$59.4\,$^\circ$C at McGrath; \citealp{Tanaka90}).  Records were also set
for the number of days below a certain temperature threshold (e.g.~6
days of less than $-$40.0\,$^\circ$C at Fairbanks; \citealp{Tanaka90}).

The period of low temperatures was caused by a large static
high-pressure system which remained over the state for two weeks
before moving southwards, breaking records in the lower 48 states as
it went \citep{Tanaka90}.  The period immediately following
this cold snap, in early February, was then much warmer than average
(by 18\,$^\circ$C for the monthly mean in Barrow).

The daily average temperatures for 1989 show this period of
exceptionally low temperatures clearly for McGrath and Fairbanks
(Fig.~\ref{fig:17}).  The traces include the short period of warming during
the middle of the cold snap which was reported in Fairbanks.  The
rapid warming and subsequent high temperatures are also detected at
both stations.  Figure~\ref{fig:17} also shows the synoptic resolution data for
January and February~1989.  These show the full extent of the cold
snap.  The minimum temperature in HadISD for this period in McGrath was
$-$58.9\,$^\circ$C (only 0.5\,$^\circ$C warmer than the new record) and $-$46.1\,$^\circ$C at
Fairbanks.  As HadISD is a sub-daily resolution dataset, then the true
minimum values are likely to have been missed, but the dataset still
captures the very cold temperatures of this event.  Some observations
over the two week \mbox{period} were flagged, from a mixture of the gap,
climatological, spike and odd cluster checks, and some were removed by
the month clean-up.  However, they do not prevent the detailed
analysis of the event.

\subsection{Australian heat waves, January \& November~2009}\label{sec:valid:australia}

South-eastern Australia experienced two heat waves during 2009.  The
first, starting in late January, lasted approximately two weeks.  The
highest temperature recorded was 48.8\,$^\circ$C in Hopetoun, Victoria, a new
state record, and Melbourne reached 46.4\,$^\circ$C, also a record for the
city.  The duration of the heat wave is shown by the record set in
Mildura, Victoria, which had 12 days where the temperature rose to
over 40\,$^\circ$C.

The second heat wave struck in mid-November, and although not as
extreme as the previous, still broke records for November
temperatures.  Only a few stations recorded maxima over 40\,$^\circ$C but many
reached over 35\,$^\circ$C.

In Fig.~\ref{fig:18} we show the average daily temperature calculated from the
HadISD data for Adelaide and Melbourne and also the full synoptic
resolution data for January and February~2009.  Although these plots
are complicated by the diurnal cycle variation, the very warm
temperatures in this period stand out as exceptional.  The maximum
temperatures recorded in the HadISD in Adelaide are 44.0\,$^\circ$C and 46.1\,$^\circ$C
in Melbourne.  The maximum temperature for Melbourne in the HadISD is
only 0.3\,$^\circ$C lower than the true maximum temperature.  However, some
observations over each of the two week periods were flagged, from a
mixture of the gap, climatological, spike and odd cluster checks, but
they do not prevent the detailed analysis of the event.

\subsection{Global overview of the quality control procedure}\label{sec:valid:globaloverview}

The overall observation flagging rates as a percentage of total number
of observations are given in Fig.~\ref{fig:19} for temperature, dewpoint
temperature and sea-level pressure. Disaggregated results for each
test and variable are summarised in Table~\ref{table:6}.  For all variables the
majority of stations have $<1$ per cent  of the total number of
observations flagged. Flagging patterns are spatially distinct for
many of the individual tests and often follow geopolitical rather than
physically plausible patterns (Table~\ref{table:6}, final column), lending
credence to a non-physical origin. For example, Mexican stations are
almost ubiquitously poor for sea-level pressure measurements. For the
three plotted variables, rejection rates are also broadly inversely
proportional to natural climate variability (Fig.~\ref{fig:19}). This is
unsurprising because it will always be easier to find an error of a
given absolute magnitude in a time series of intrinsically lower
variability. From these analyses we contend that the QC procedure is
adequate and unlikely to be over-aggressive.

In a number of cases, stations that had apparently high flagging
rates for certain tests were also composite stations (see figures for
the tests).  In order to check whether the compositing has caused more
problems than it solved, 20 composite stations were selected at random
to see if there were any obvious discontinuities across their entire
record using the raw, non-quality-controlled data.  No such problems were found in
these 20 stations.  Secondly, we compared the flagging prevalence (as
in Table~\ref{table:9}) for each of the different tests focussing on the three
main variables.  For most tests the difference in flagging percentages
between composite and non-composite stations is small.  The most
common change is that there are fewer composite stations with 0 per
cent of data flagged and more stations with 0--0.1 per cent of data
flagged than non-composites. We do not believe these differences
substantiate any concern. However, there are some cases of note. In
the case of the dewpoint cut-off test, there is a large tail out to
higher failure fractions, with a correspondingly much smaller 0 per
cent flagging rate in the case of composite stations.  There is a
reduction in the prevalence of stations which have high flagging rates
in the isolated odd cluster test in the composite stations versus the
non-composite stations.  The number of flagging due to streaks of all
types is elevated in the composite stations.

Despite no pervasive large differences being found in apparent data
quality between composited stations and non-composited stations, there
are likely to be some isolated cases where the compositing has caused
a degrading of the data quality. Should any issues become apparent to
the user, feedback to the authors is strongly encouraged so that
amendments can be made where possible.

%t7
\begin{table*}[t]
\caption{\label{table:7} Data precision and reporting
  interval by month for all of the 6103 stations (\textit{.all}) and the 3427
  filtered stations (\textit{.clim}).  Months with no data at all are not
  counted, but those with few data are unlikely to have well-determined accuracies or reporting intervals and will fall under the
unable-to-identify category.}
\begin{tabular}{lrrrrrr}
\tophline
&\multicolumn{2}{c}{Temperature}&\multicolumn{2}{c}{Dewpoint}&\multicolumn{2}{c}{SLP}\\
\hhline
&.all       &  .clim    &.all       &.clim          &.all     &.clim      \\
\hhline
\multicolumn{7}{c}{Data Precision}\\
\hhline
Unable to identify      &2.70\,\%     &  0.80\,\%   &3.60\,\%        &1.20\,\%      &27.90\,\%  &20.10\,\%     \\
0.1                     &49.70\,\%    &  51.10\,\%  &50.50\,\%       &51.70\,\%     &71.10\,\%  &78.70\,\%     \\
0.5                     &2.00\,\%     &  1.20\,\%   &0.30\,\%        &0.30\,\%      &0.30\,\%   &0.40\,\%      \\
1                       &45.50\,\%    &  46.80\,\%  &45.50\,\%       &46.80\,\%     &0.80\,\%   &0.70\,\%      \\
\hhline
\multicolumn{7}{c}{Reporting Interval (hours)}\\
\hhline
Unable to identify      &4.80\,\%     &  1.80\,\%   &5.80\,\%        &2.30\,\%      &29.50\,\%  &21.30\,\%     \\
1                       &31.00\,\%    &  36.80\,\%  &30.60\,\%       &36.60\,\%     &28.60\,\%  &39.90\,\%     \\
2                       &4.20\,\%     &  1.70\,\%   &4.00\,\%        &1.60\,\%      &3.10\,\%   &1.70\,\%      \\
3                       &59.80\,\%    &  59.60\,\%  &59.40\,\%       &59.50\,\%     &38.60\,\%  &37.00\,\%     \\
4                       &0.30\,\%     &  0.10\,\%   &0.30\,\%        &0.10\,\%      &0.30\,\%   &0.10\,\%      \\
\bottomhline
\end{tabular}\vspace{3mm}
\end{table*}

The data recording resolution (0.1, 0.5 or whole number) and
reporting intervals (1-, 2-, 3- and 4-hourly.) summarised over all stations
in HadISD are in Table~\ref{table:7}.  There is a clear split in the
temperature and dewpoint data resolution between whole degrees
and 1/10th degree.  Most of the sea-level pressure measurements are to
the nearest 1/10th of a hPa.  These patterns are even stronger when
using only the 3427 \textit{.clim} stations (see Sect.~6).  The
reporting intervals are  mostly at hourly-
and three-hourly intervals, and rarely at two- or four-hourly
intervals.  The reporting interval could not be determined in a
comparatively much larger fraction of sea-level pressure observations
than in temperature or dewpoint.

\section{Final station selection}\label{sec:finalselection}

Different end-users will have different requirements for data
completeness and quality. All records passing QC are available in
HadISD versions ``\textit{.all}'', but further checks are performed on stations for
inclusion in HadISD versions ``\textit{.clim}'', to ensure adequacy for
long-term climate monitoring. These
additional checks specify a minimum temporal completeness and quality criteria
using three categories: temporal record completeness; reporting
frequency; and proportion of values flagged during QC. All choices
made here are subjective, and parameters could arguably be changed
depending on desired end-use. Table~\ref{table:8} summarises the thresholds used
here for station inclusion. The final network composition results in
3427 stations and is given in Fig.~\ref{fig:20} which also shows the stations
that were rejected and which of the station inclusion criteria of
individual stations are rejected and why.

%t8
\begin{table*}[t]
\caption{\label{table:8} Station inclusion criteria: ranges considered
and final choices.  Note that there has been no selection on the wind or
 cloud variables.  These variables have not been the focus of the QC
  procedure; we therefore do not exclude stations which have valid
  temperature, dewpoint and pressure data on the basis of their wind and cloud data quality.}
\begin{tabular}{p{4.5cm}ll}
\tophline
Parameter&Range considered          &Final choice     \\
\hhline
\multicolumn{3}{c}{Record completeness} \\
\hhline
First data point before&1 January 1975--1 January 1990        &1 January 1980      \\
Last data point after&31 December 1990--31 December 2005        &31 December 2000      \\
\hhline
\multicolumn{3}{c}{Temporal completeness}\\
\hhline
Quartiles of diurnal cycle sampled for day to count& \multicolumn{1}{c}{2--4}                &\multicolumn{1}{c}{3}          \\
Days in month for month to count&\multicolumn{1}{c}{12, 20, 28 }            &\multicolumn{1}{c}{12 }        \\
Years for a given calendar month present to count as complete&\multicolumn{1}{c}{10, 15, 20, 25, 30}        &\multicolumn{1}{c}{20}         \\
Number of months passing completeness criteria for year to count&\multicolumn{1}{c}{9, 10, 11, 12}               &\multicolumn{1}{c}{10 }        \\
Maximum continuous gap&\multicolumn{1}{c}{0, 1, 2, 3, 4\,yr}        &\multicolumn{1}{c}{2\,yr}   \\
\hhline
\multicolumn{3}{c}{Reporting frequency} \\
\hhline
Median reporting time interval&\multicolumn{1}{c}{1, 3, 6\,h}                  &\multicolumn{1}{c}{3\,h}   \\
\hhline
\multicolumn{3}{c}{Quality control (all tests applied only if more than 20\,\% of time steps report this variable)}\\
\hhline
T QC flag prevalence&\multicolumn{1}{c}{1, 2, 5, 10\,\%}                 & \multicolumn{1}{c}{$<5$\,\%}          \\
Td QC flag prevalence&\multicolumn{1}{c}{1, 2, 5, 10\,\%}                & \multicolumn{1}{c}{$<5$\,\%}          \\
SLP QC flag prevalence&\multicolumn{1}{c}{1, 2, 5, 10\,\%}               & \multicolumn{1}{c}{$<5$\,\%}          \\
ws QC flag prevalence&\multicolumn{1}{c}{10, 20, 100\,\%}               & \multicolumn{1}{c}{$<100$\,\%}        \\
wd QC flag prevalence&\multicolumn{1}{c}{10, 20, 100\,\% }              & \multicolumn{1}{c}{$<100$\,\%}        \\
Cloud total QC flag prevalence&\multicolumn{1}{c}{50, 100\,\%}               & \multicolumn{1}{c}{$<100$\,\%}        \\
High cloud QC flag prevalence&\multicolumn{1}{c}{50, 100\,\%}                & \multicolumn{1}{c}{$<100$\,\%}        \\
Medium cloud QC flag prevalence&\multicolumn{1}{c}{50, 100\,\%}              & \multicolumn{1}{c}{$<100$\,\%}        \\
Low cloud QC flag prevalence&\multicolumn{1}{c}{50, 100\,\%}                 & \multicolumn{1}{c}{$<100$\,\%}        \\
\bottomhline
\end{tabular}\vspace{3mm}
\end{table*}

%t9
\begin{table*}[t]
\caption{\label{table:9}As Table~\ref{table:6} but in percentages.
  Summary of removal of data from individual stations by each test for
  the 6103 stations in the .all dataset.  Each row shows the
  percentage of stations that had fractional removal rates in the
  seven bands for the test and variable indicated.}
{
\begin{tabular}{p{95pt}p{40pt}llllllll}
\tophline
Test&Variable&\multicolumn{8}{c}{Stations within detection rate band (\% of total original observations)}\\
\cline{3-10}
&&0       &0--0.1    &0.1--0.2      &0.2--0.5      &0.5--1.0       &1.0--2.0  &2.0--5.0      &$>5.0$   \\
\hhline
Duplicate months data&All&100.0     &0.0        &0.0     &0.0     &0.0     &0.0      &0.0     &0.0      \\
\hhline
Isolated cluster&T&33.4    &45.7     &7.9   &6.8   &3.5    &2.1    &0.6   &0.0      \\
&Td&30.4    &48.3     &7.9   &7.2   &3.8    &2.1    &0.6   &0.0     \\
&SLP&26      &51.6     &9.3   &8.0     &3.3    &1.5    &0.3   &0.0      \\
&ws&32.1    &46.7     &7.9   &7.1   &3.6    &2.0      &0.5   &0.1    \\
\hhline
Frequent values&T&98.0      &1.5      &0.2   &0.1   &0.1    &0.0      &0.0     &0.0      \\
&Td&97.3    &1.5      &0.3   &0.3   &0.2    &0.1    &0.1   &0.1    \\
&SLP&98.3    &0.4      &0.1   &0.1   &0.1    &0.1    &0.5   &0.3    \\
\hhline
Diurnal cycle&All&94.5    &0        &0.3   &2.9   &1.1    &0.6    &0.4   &0.2    \\
\hhline
Distributional gap&T&42.1    &53.3     &0.7   &1.3   &1.3    &0.5    &0.6   &0.1    \\
&Td&18.9    &68.9     &4.9   &4.0     &1.9    &0.7    &0.6   &0.1    \\
&SLP&44.8    &50.7     &1.2   &1.5   &0.9    &0.5    &0.3   &0.1    \\
\hhline
Record check&T&87.1    &12.9     &0.0     &0.1   &0.0      &0.0      &0.0     &0.0      \\
&Td&99.8    &0.2      &0.0     &0.0    &0.0      &0.0     &0.0     &0.0      \\
&SLP&79.9    &20.1     &0.0     &0.0     &0.0     &0.0      &0.0     &0.0      \\
&ws&100.0     &0.0        &0.0     &0.0     &0.0      &0.0      &0.0     &0.0      \\
\hhline
Repeated streaks/unusual spell frequency&T&78.4    &4.2      &3.0     &4.7   &3.6    &3.5    &2.4   &0.3    \\
&Td&71.2    &3.6      &3.1   &5.7   &5.6    &5.8    &4.5   &0.5    \\
&SLP&98.2    &0.4      &0.2   &0.2   &0.1    &0.1    &0.4   &0.3    \\
&ws&88.7    &4.0        &2.4   &2.2   &1.2    &1.0      &0.4   &0.1    \\
\hhline
Climatological outliers&T&21.2    &71.8     &3.6   &2.6   &0.6    &0.2    &0.0     &0.0      \\
&Td&17.4    &74.4     &3.9   &3.1   &0.8    &0.3    &0.0     &0.0      \\
\hhline
Spike check&T&1.6     &59.8     &20.8  &16.3  &1.5    &0.0      &0.0     &0.0      \\
&Td&0.6     &58.4     &24.3  &15.4  &1.1    &0.1    &0.0     &0.0      \\
&SLP&12.5    &56.3     &17.5  &13.1  &0.5    &0.0      &0.0     &0.0      \\
\hhline
Supersaturation&T, Td&100.0     &0.0        &0.0     &0.0     &0.0      &0.0      &0.0     &0.0      \\
\hhline
Wet bulb drying&Td&65.2    &28.2     &3.2   &2.3   &0.6    &0.4    &0.1   &0.0      \\
\hhline
Wet bulb cutoffs&Td&82.8    &1.9      &3.5   &5.2   &2.9    &2.1    &1.1   &0.5    \\
\hhline
Cloud clean-up&Cloud \par variables            &0.0       &11.2     &7.7   &16.0    &18.4   &22.2   &19.2  &5.2    \\
\hhline
Unusual variance&T&92.7    &0.2      &1.3   &4.9   &0.8    &0.1    &0.0     &0.0      \\
&Td&91.8    &0.2      &1.9   &5.5   &1.0      &0.2    &0.0     &0.0      \\
&SLP&86.2    &0.5      &1.9   &8.1   &2.5    &0.4    &0.2   &0.1    \\
\hhline
Neighbour differences&T&25.4    &71.6     &4.5   &0.6   &0.4    &0.3    &0.2   &0.0      \\
&Td&23.9    &71.6     &2.6   &1.0     &0.6    &0.3    &0.2   &0.0      \\
&SLP&29.9    &65.5     &3.3   &1.0     &0.2    &0.1    &0.1   &0.0      \\
\hhline
Station clean-up&T&63.3    &25.3     &3.9   &3.6   &2.3    &1.0      &0.4   &0.1    \\
&Td&61.5    &25       &3.9   &4.5   &2.6    &1.4    &0.8   &0.2    \\
&SLP&53.2    &36.7     &3.5   &3.7   &1.8    &0.7    &0.5  &0.1    \\
&ws&63.9  &26.0        &3.5  &3.7   &1.8    &0.7  &0.4   & 0.1\\
\tophline
\end{tabular}}
\vspace{3mm}
\end{table*}

The huge majority of rejected stations fail on record completeness
(1234) or because the first (last) observation occurs too late (early)
(689).  Large gaps in the data
cause a further 626 stations to fail. In some regions this leads to
almost complete removal of country records (e.g.~eastern Germany,
parts of the Balkan region, Iran, Central Africa). This may be linked
to known changes in WMO station IDs for a number of countries
including renumbering countries from the former Yugoslavia
\citep{Jones03}. Record completeness rejections were insensitive
to a variety of temporal criteria (Table~\ref{table:8}), which
therefore cannot be stretched to accept more stations without
unreasonably including records that are too incomplete for end-users.  Remaining
rejections were based upon not retaining sufficient data post-QC for
one or more variables. There is a degree of clustering here with major
removals in Mexico (largely due to SLP issues), NE North America,
Alaska, the Pacific coast and Finland.

\section{Dataset nomenclature, version control and source\\ code transparency}\label{sec:nomenclature}

The official name of the dataset created herein is HadISD.1.0.0.2011f.
Within this there are two versions available: HadISD.1.0.0.2011f.all
for all of the 6103 quality-controlled stations and
HadISD.1.0.0.2011f.clim for those 3427 stations which match the above
selection criteria. Future versions will be made available that will
include new data (more stations and/or updated temporal coverage) or a
minor code change/bug fix.  An update of the data to the next calendar
year (e.g.~to 1~January~2013, 00:00\,UT) will result in the year label
incrementing to 2012.  f indicates a final dataset, whereas
other letters could indicate other specifications, for example p\,=\,preliminary.  Any updates or
changes will be described on the website or in a readme file along
with a version number change (e.g.~HadISD.1.0.1), or if considered
more major, as a technical note (e.g.~HadISD.1.1.0) depending on the
level of the change.  A major new version (e.g.~HadISD.2.0.0) will be
described in a peer-reviewed publication. The full version number is
in the metadata of each netCDF file.  Suffixes such as ``\textit{.all}''
and ``\textit{.clim}''
identify the type of dataset. These may later include new derived
products with alternative suffixes. Through this nomenclature, a user
should be clear about which version they are using. All major versions
will be frozen prior to update and archived. However, minor changes
will only be kept for the duration of the major version being live.

The source code used to create HadISD.1.0.0 is written in IDL. It will
be made available alongside the dataset at
\url{http://www.metoffice.gov.uk/hadobs/hadisd}. Users are welcome to copy
and use this code. There is no support service for this code, but
feedback is appreciated and welcomed through a comment box on the
website or by contacting the authors directly.

\section{Brief illustration of potential uses}\label{sec:uses}

Below we give two examples, highlighting the potential unique
capabilities of this sub-daily dataset in
comparison to monthly or daily holdings.

\subsection{Median diurnal temperature range}\label{sec:uses:dtr}

\vspace{-2mm}

In Fig.~\ref{fig:21} we show the median diurnal temperature range (DTR) from
the subset of 3427 \textit{.clim} stations which have records commencing before
1975 and ending after 2005 for the four standard three-month seasons.
The DTR was calculated for each day from the maximum-minimum recorded
temperature in each 24\,h period, with the proviso that there are
at least four observations in a 24\,h period, spanning at least 12\,h.

The highest DTRs are observed in arid or high altitude regions, as
would be expected given the lack of water vapour to act as a
moderating influence.  The
stark contrast between high- and low-lying regions can be seen in
Yunnan province in the south-west of China as the DTRs increase with
the station altitude to the west.

The differences between the four figures are most obvious in regions
which have high station densities, and between DJF and JJA.  The
increase in DTR associated with the summer months in Europe and
central Asia is clear.  This is coupled with a decrease in the DTR in
the Indian subcontinent and in sub-Saharan West Africa, linked to the
monsoon cycle.  Although the DJF DTR in North America is larger than
that in Europe, there is still an increase associated with the summer
months.   Stations in desert regions, e.g. Egypt and the interior of
Australia, as well as those in tropical maritime climates show very
consistent DTRs in all seasons.

\vspace{-2mm}

\subsection{Temperature variations over 24 hours}\label{sec:uses:tempvar}

\vspace{-2mm}

In Fig.~\ref{fig:22} we show the station temperature from all the 6103
stations in the \textit{.all} dataset over the entire the globe, which pass the
QC criteria, for 00:00, 06:00, 12:00 and 18:00\,UT on 28~June~2003.
The evolution of the highest temperatures with longitude is as would
be expected.  The highest temperatures are also seen north of the
Equator, as would be expected for this time of year.  Coastal stations at high
latitudes show very little change in the temperatures,
and those in Antarctica especially so, as it is the middle of their
winter.  In the lower two panels the lag of the location of the
maximum temperature behind the local midday can be seen.  At 12:00\,UT,
the maximum temperatures are still being experienced in Iran and the
surrounding regions, and at 18:00\,UT, they are seen in northern and
western sub-Saharan Africa.  We note the one outlier in Western Canada
at 18:00\,UT, which has been missed by the QC suite.

\vspace{-2mm}

\conclusions[Summary]\label{sec:summary}

\vspace{-2mm}

We have developed a long-term station
subset, HadISD, of the very large ISD synoptic report database
\citep{Smith11}, in a more scientific analysis, user-friendly {netCDF} data format
together with an alternative quality control suite to better span
uncertainties inherent in quality control procedures. We note that
the raw ISD data may have differing \mbox{levels} of QC applied by National
Met Services before ingestion into the ISD. For HadISD, assigned duplicate
stations were composited. The data were then converted to netCDF
format for those stations with plausibly climate-applicable
record characteristics.  Intra- and inter-station quality control
procedures were developed and refined with reference to a small subset
of the network and a limited number of UK-based case studies. Quality
control was undertaken on temperature, dewpoint temperature, sea-level
pressure, winds, and clouds, focusing on the first three, to which
highest confidence can be attached. Quality controls were sequenced so
that the worst data were flagged by earlier tests and
subsequent tests became progressively more sensitive. Typically less
than 1 per cent of the raw synoptic data were flagged in an individual
station record.  Finally, we applied selection criteria based upon
record completeness and QC flag indicator frequency, to yield a final
set of stations which are recommended as suitable for climate
applications. A few case studies were used to confirm the efficacy of
the quality control procedures
and illustrate some potential simple applications of HadISD.  The
dataset has a wide range of applications, from the study of individual
extreme events to the change in the frequency or severity of these
events over the span of the data, the results of which can be compared
to estimates of past extreme events and those in projected future
climates.

The final dataset (and an audit trail) is available on
\url{http://www.metoffice.gov.uk/hadobs/hadisd} for bona fide research
purposes and consists of over 6\,000 individual station records from
1973 to 2011 with near global coverage (\textit{.all}) and over 3400 stations
with long-term climate quality records (\textit{.clim}).  A version
control and archiving
system has been created to enable the clear identification of which
version of HadISD is being used, along with any future changes from
the \mbox{methodology} outlined herein.

\section*{Copyright statement}

This work is distributed under the Creative Commons Attribution 3.0
License together with an author copyright. This license does not conflict
with the regulations of the Crown Copyright.

\begin{acknowledgements}
We thank Neal Lott and two anonymous referees for their useful and
detailed reviews which helped improve the final manuscript and dataset.

The Met Office Hadley Centre authors were supported by the Joint
DECC/Defra Met Office Hadley Centre Climate Programme (GA01101). Much
of P.~W.~Thorne's early effort was supported by NCDC, and the Met
Office PHEATS contract. The National Center for Atmospheric Research
is sponsored by the US National Science Foundation. E.~V.~Woolley
undertook work as part of the Met Office summer student placement
scheme whilst an undergraduate at Exeter University. We thank Peter
Olsson (AEFF, UAA) for assistance.\\
\\
Edited by: H.~Goosse
\end{acknowledgements}

\bibliography{hadisd_arxiv}
\bibliographystyle{plainnat}

\end{document}